\documentclass[sigconf]{acmart}
\AtBeginDocument{%
  }

\setcopyright{acmlicensed}
\copyrightyear{2018}
\acmYear{2018}
\acmDOI{XXXXXXX.XXXXXXX}
\acmConference[Conference acronym 'XX]{Make sure to enter the correct
  conference title from your rights confirmation email}{June 03--05,
  2018}{Woodstock, NY}
\acmISBN{978-1-4503-XXXX-X/2018/06}
\usepackage{algorithm}
\usepackage{algpseudocode}

\newcommand{\Oursystem}{Canvas3D}

\begin{document}

\title{Canvas3D: Empowering Precise Spatial Control for Image Generation with Constraints from a 3D Virtual Canvas}

\author{Runlin Duan}
\orcid{0000-0001-8256-6419}
\authornote{Both authors contributed equally to this research.}
\affiliation{%
  \institution{School of Mechanical Engineering \\ Purdue University}
  \city{West Lafayette}
  \country{USA}}
\email{duan92@purdue.edu}

\author{Yuzhao Chen}
\orcid{0009-0005-6196-1176}
\authornotemark[1]
\affiliation{%
  \institution{Elmore Family School of Electrical and Computer Engineering \\ Purdue University}
  \streetaddress{610 Purdue Mall}
  \city{West Lafayette}
  \state{IN}
  \country{USA}
  \postcode{47907}
}
\email{chen4863@purdue.edu}

\author{Rahul Jain}
\affiliation{%
  \institution{Elmore Family School of Electrical and Computer Engineering \\ Purdue University}
  \city{West Lafayette}
  \country{USA}}
\email{jain348@purdue.edu}

\author{Yichen Hu}
\orcid{0009-0001-9385-4456}
\affiliation{%
  \institution{Department of Computer Science}
  \city{West Lafayette}
  \country{USA}}
\email{hu925@purdue.edu}

\author{Jingyu Shi}
\orcid{0000000151592235}
\affiliation{%
  \institution{Elmore Family School of Electrical and Computer Engineering \\ Purdue University}
  \city{West Lafayette}
  \country{USA}}
\email{shi537@purdue.edu}

\author{Karthik Ramani}
\orcid{0000-0001-8639-5135}
\affiliation{%
  \institution{School of Mechanical Engineering \\ Purdue University}
  \city{West Lafayette}
  \country{USA}}
\email{ramani@purdue.edu}



\begin{abstract}
Generative AI (GenAI) has significantly advanced the ease and flexibility of image creation. 
However, it remains a challenge to precisely control spatial compositions, including object arrangement and scene conditions. 
To bridge this gap, we propose \Oursystem{}, an interactive system leveraging a 3D engine to enable precise spatial manipulation for image generation. 
Upon user prompt, \Oursystem{} automatically converts textual descriptions into interactive objects within a 3D engine-driven virtual canvas, empowering direct and precise spatial configuration. 
These user-defined arrangements generate explicit spatial constraints that guide generative models in accurately reflecting user intentions in the resulting images.
We conducted a closed-end comparative study between \Oursystem{} and a baseline system. 
And an open-ended study to evaluate our system "in the wild".
The result indicates that \Oursystem{} outperforms the baseline on spatial control, interactivity, and overall user experience.

\end{abstract}

\ccsdesc[500]{Do Not Use This Code~Generate the Correct Terms for Your Paper}
\ccsdesc[300]{Do Not Use This Code~Generate the Correct Terms for Your Paper}
\ccsdesc{Do Not Use This Code~Generate the Correct Terms for Your Paper}
\ccsdesc[100]{Do Not Use This Code~Generate the Correct Terms for Your Paper}

\keywords{Controllable Image Generation, Spatial Control, Conditional Generative Models}
\begin{teaserfigure}
  \includegraphics[width=\textwidth]{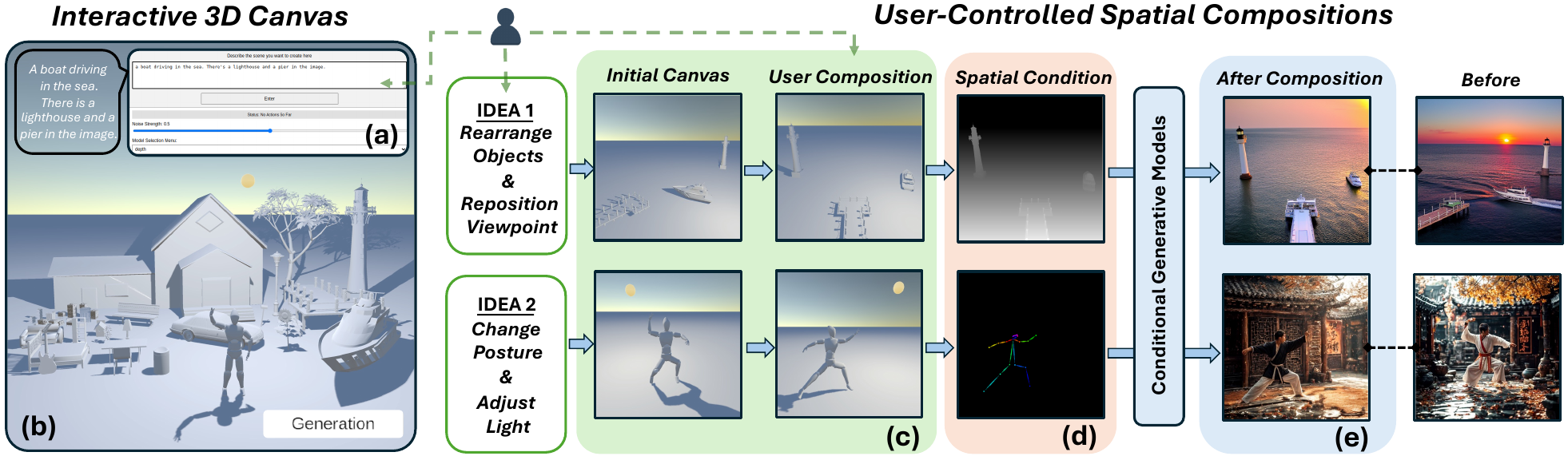}
  \caption{The workflow of controlling spatial composition using \Oursystem{}. A user starts by prompting the desired contents to the system (a). The system parses this prompt to create an interactive 3D environment, referred to as "canvas" (b). The user then interacts with the canvas to specify the desired composition of objects and environmental factors (c). In one case, the user first rearranges the placement of the boat and the lighthouse, and then changes the viewpoint. In another case, the user repositions the human pose and changes the lighting direction from left to right. The system translates the user-specified arrangement into explicit spatial constraints, referred to as "spatial conditions" for generative models (d). Finally, these conditions guide the generative model to produce images that accurately reflect the user's spatial intents (e).}
  \Description{}
  \label{fig:teaser}
\end{teaserfigure}


\maketitle

\section{Introduction}

Generative AI (GenAI) has emerged as a transformative technique for image generation \cite{rombach2022high,ramesh2021zero,ramesh2022hierarchical}.
This new technology enables users to easily create images on demand and rapidly explore different compositions of image features, such as color \cite{hou2024c2ideas}, style \cite{oh2024lumimood}, and object shapes \cite{lin2024inkspire,sarukkai2024block}. 
However, there remains a challenge to precisely control the spatial composition of images \cite{zhang2023adding}.
For example, text-to-image generation models do not properly capture spatial constraints, such as object placement and the relationship between different objects \cite{huang2025t2i, sun2023dreamsync}. 

Imagine a scenario in which a user wants to generate a room with a particular arrangement of furniture and decor, then visualize it from various perspectives. 
To control this image, the user must be able to precisely describe their spatial intents, such as furniture placement, to the generative model.
Prior research has explored various methods to collect user spatial intents, such as text \cite{brade2023promptify,choi2024creativeconnect}, sketches \cite{sarukkai2024block,lin2024inkspire}, and dragging \cite{shi2024dragdiffusion,avrahami2024diffuhaul}.
However, these approaches depend on low-dimensional modalities to describe high-dimensional spatial compositions, which lack precision or demand significant user effort. 
Specifically, text-based methods, while intuitive, are typically imprecise and require extensive trial-and-error processes \cite{brade2023promptify}. 
In addition, sketch-based methods can provide precise spatial guidance but require substantial sketching skills, spatial awareness, and visualization abilities from users \cite{sarukkai2024block}. 
Meanwhile, dragging-based methods \cite{pan2023drag,shin2024instantdrag,shi2024dragdiffusion}, simplify spatial edits via direct 2D manipulation but lack depth manipulation capabilities \cite{avrahami2024diffuhaul}, which result in objects failing to blend into new positions naturally.

On the other hand, precisely generating user-defined spatial composition requires the generative model to accurately leverage user spatial constraints.
Recent research introduces conditional generative models complying with precise 3D spatial constraints, e.g., depth maps \cite{zhang2023adding} and human skeletons \cite{li2024simple}.
However, initializing such spatial constraints for conditional generative models poses significant challenges for out-of-lab use.
For instance, current approaches commonly rely on sliders to specify the position and rotation of each object to initialize the depth condition \cite{bhat2024loosecontrol,eldesokey2024build}. 
The slider-based interaction causes strain due to the attention switch \cite{visser1999attentional,gopher2000switching}-users have to switch their attention back and forth between the control widgets and the object states,  which is tedious and impractical for compositing multiple objects. 
Given the challenge with existing approaches, we aim to improve the alignment between user intent and generated image by developing an intuitive interaction mechanism that (1) accurately captures user spatial intents, and (2) seamlessly translates user intents into inputs that generative models can understand and execute.

To address these considerations, we turn our attention to 3D engines \cite{UnityWebsite}, which have proven highly precise in other domains for native spatial manipulation \cite{qian2022arnnotate,autodesk2025,ros2025}. 
The 3D engine simulates the object and renders the environment in a virtual 3D space.
It offers natural and intuitive spatial interaction by empowering users to directly manipulate objects using common input modalities \cite{jankowski2013survey,jankowski2013survey}, such as a mouse and keyboard \cite{schultheis2012comparison}.
By adopting 3D engine-driven interactions, we can develop a system that allows precise spatial control by supporting direct spatial manipulations in a 3D environment.

We propose \Oursystem{}, an image generation system with precise spatial control.
When using \Oursystem{}, the user first provides a textual prompt describing desired content (Figure 1-a); the system parses this description into an interactive 3D environment, referred to as "canvas" (Figure 1-b).
Users then directly manipulate the objects in the canvas to specify the desired spatial composition (Figure 1-c).
The completed object arrangements automatically translate into explicit spatial constraints, referred to as "spatial conditions",  for generative models (Figure 1-d).
These conditions guide generative models to produce images that accurately reflect user intents (Figure 1-e).

We highlight our contributions as follows:
\begin{itemize}
    \item A systematic workflow empowering users to generate images with precise spatial control.
    \item An integrated pipeline that automatically translates textual prompts into an interactive 3D canvas.
    \item A unified user interface combining spatial manipulation with run-time image generation. 
    \item Two user studies evaluating system usability and demonstrating enhanced spatial controllability compared to a baseline system.
\end{itemize}

\section{Related Work}

\subsection{Interaction Design for Image Generation}

The emergence of recent generative models has inspired research to explore their flexible image-generation capabilities in a wide range of tasks, including design\cite{wadinambiarachchi2024effects, wang2024roomdreaming}, ideation\cite{jeon2021fashionq, koch2019may}, and self-expression\cite{shi2024personalizing}.
Through these prior explorations, research has highlighted the importance of controlling generated images to meet the specific needs in various domains.
For instance, in product design, researchers might explore various combinations of functionality or structure\cite{liu20233dall}, whereas in interior design, the research focus may lean toward consistency in style and color\cite{hou2024c2ideas}.

Fine-grained control over image generation also allows everyday users to tailor images to their needs and aesthetic taste, such as adjusting colors\cite{wang2023language}, moods\cite{yang2024emogen,oh2024lumimood}, or arranging elements to fit a particular aesthetic \cite{wang2024roomdreaming}. 
These approaches lower the barriers for image creation and make it an approachable activity.
However, spatial control has been largely overlooked, despite the existing research emphasising aesthetic expression.
For instance, while many studies have explored how to manipulate color and style, few have addressed the challenges of configuring spatial composition in generated images.
Specifically, users can not precisely specify desired layouts, character poses, lighting conditions, and perspectives to generate images that accurately match their mental imagery.

Previous controllable image generation systems typically manage spatial composition through lower-dimensional input modalities, such as text prompts\cite{liao2022text}, sketches\cite{sarukkai2024block, lin2025inkspire}, or dragging \cite{dang2022ganslider,shi2024dragdiffusion,avrahami2024diffuhaul}.
However, these systems often face limitations, including imprecision or the requirement of additional user skills when attempting to articulate complex three-dimensional compositions using simpler input methods.
For example, text prompts do not properly capture spatial constraints, such as object placement and the relationship between different objects \cite{huang2025t2i, sun2023dreamsync}.
Using sketches to precisely express spatial arrangements requires practice \cite{williford2019framework}.
Dragging control, although user-friendly, cannot adequately handle depth manipulation \cite{avrahami2024diffuhaul}, which results in objects failing to blend into new positions naturally.

These limitations highlight a critical research opportunity: developing an accurate spatial control method that empowers effortless user intent expression.
\Oursystem{} bridges this gap by capturing the user intent through collecting the explicit arrangement of objects and the environment in 3D space.

\subsection{Controllable Geneartive Models}

Spatial control has been a focus in creating controllable generative models \cite{gokhale2022benchmarking, liao2022text}.
Inspired by traditional computer vision research on depth estimation and contour detection, recent researchers \cite{zheng2023layoutdiffusion, mou2024t2i, zhang2023adding} have proposed conditional generative models that leverage conditional images containing accurate spatial information.
To improve the spatial accuracy, recent conditional generative models accept a text prompt alongside conditional inputs, such as depth images, to provide supplementary spatial context \cite{qin2023unicontrol, li2024simple}.

Building on this concept, researchers have further explored the spatial conditions to encompass a broad range of spatial features, including illumination\cite{zhang2025scaling}, posture\cite{ju2023humansd}, object relationships\cite{chefer2023attend}, and numerical accuracy\cite{feng2023layoutgpt}.
These conditional generative models are compatible with various data formats that encode spatial information for precise control. 
For example, depth images represent precise object placement, human skeleton maps capture the specific human pose and movement, and object relation graphs illustrate the interaction between objects.
However, despite the improved spatial awareness of contemporary generative models, creating precise conditions to reflect users' spatial intents remains challenging for end-users.
For instance, recent approaches\cite{eldesokey2024build, bhat2024loosecontrol} require users to adjust multiple sliders to control the placement, orientation, and scales of objects.
Such cumbersome interaction designs are impractical for the out-of-lab use of everyday users.

\Oursystem{} streamlines the process to create spatial conditions by automatically encoding the conditions from the native 3D arrangement specified by users.

\subsection{Spatial Control using 3D Engine}

Prior works adopt the spatial manipulation of real-world objects to control image generation and virtual content creation. 
For example, users can assemble real-world objects to create a reference for product ideation \cite{zhang2024protodreamer} or adopt everyday objects for creating animations \cite{li2024anicraft}.
However, users may not always find suitable real-world objects to serve as references for controlling the generation.

3D engines support object manipulation by simulating objects and contextual factors within a virtual environment \cite{eberly20063d}. 
Through object and environment parametrization, 3D engines enable users to conduct spatial manipulation on objects in an unconstrained manner, overcoming limitations imposed by real-world environments \cite{jankowski2013survey}.
The precision of spatial control supported by this technology has inspired researchers to develop systems for various applications, including hand-object data collection\cite{qian2022arnnotate}, robotic training \cite{degrave2019differentiable}, modeling\cite{duan2025parametric}, and gaming\cite{neroni2021virtual}.
Additionally, 3D engines provide real-time physical constraints during spatial manipulation, ensuring that user-generated spatial compositions align with physical laws. 

In this way, adopting 3D engines for spatial control not only accurately captures users' spatial intents but inherently resolves issues on physics violations in image generation \cite{meng2025grounding, liu2025generative}.
Additionally, because of the parametrization, the spatial information generated by 3D engines can be explicitly translated into conditions compatible with generative models.
Building upon these advantages, we embrace a 3D engine to construct our image generation system that (1) intuitively and losslessly captures human spatial intents, and (2) accurately translates these intents into precise input conditions for generative models.

\section{Interaction Design for Canvas3D}

\subsection{Taxonomy of Spatial Composition}

To support the proposed spatial control method, we first establish a taxonomy of factors that define how spatial compositions are modeled, controlled, and interpreted within our system.
As described by physiologists, "Space is defined only by what encloses it." \cite{dolins2010spatial}, spatial composition rooted in the human perceptions of the surrounding environment.
It represents the arrangement and presentation of objects and contextual elements in the physical world \cite{gibson2014ecological,connor2017integration,galati2010multiple}.
For instance, when navigating a street, we perceive objects such as cars, trees, and pavements as well as their spatial relationships to guide us.

\begin{figure}[h]
  \centering
  \includegraphics[width=\linewidth]{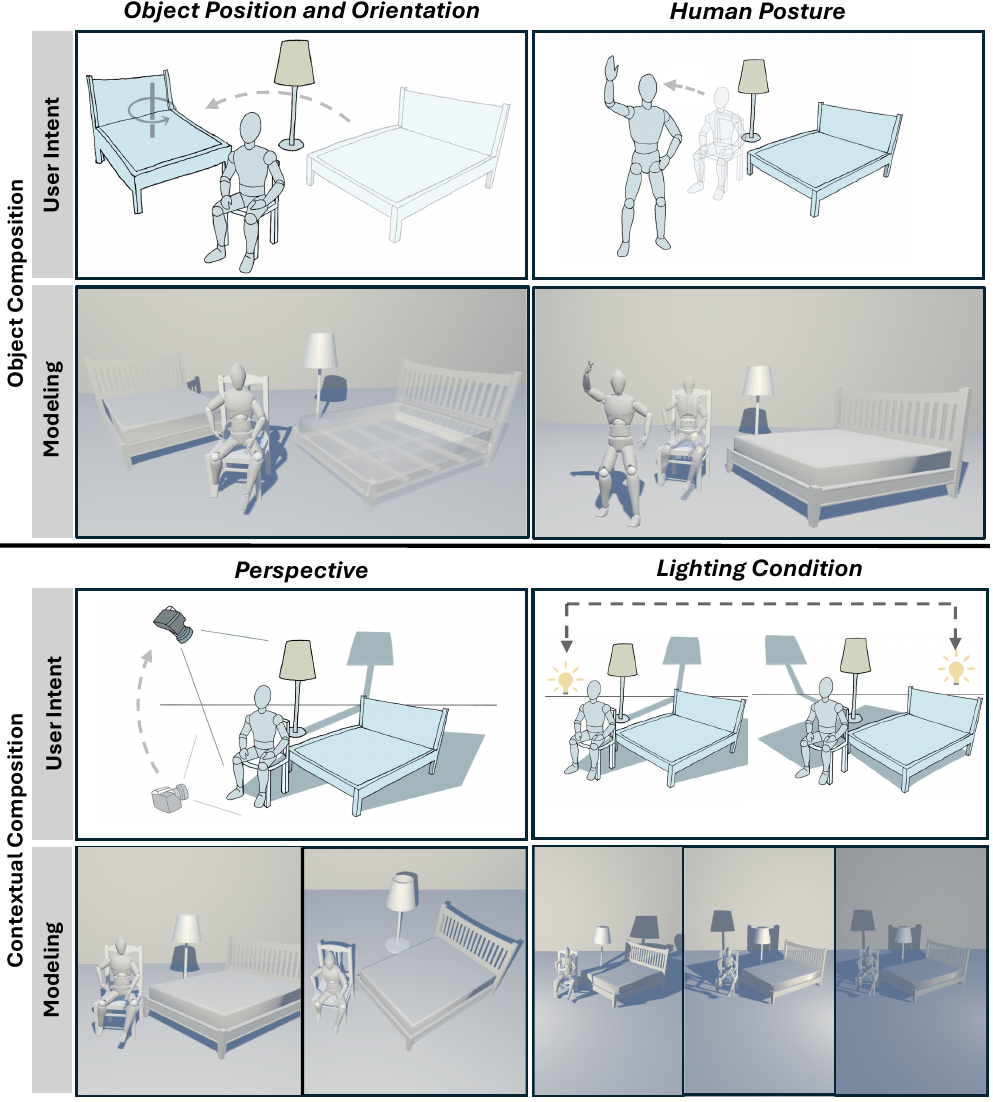}
  \caption{The taxonomy of spatial composition. Including the object composition and the contextual composition. The object composition depicts the arrangement of objects and the configuration of human posture. The contextual composition depicts the configuration of lighting and the image perspective.}
  \Description{None}
 \label{fig:taxonomy}
\end{figure}

In 2D visual design, such as painting, photography, and image generation, controlling the spatial composition enables creators to convey precise information and express desired impressions \cite{palmer2013visual,fairbrother1974nature}. 
Prior visual design studies have systematically explored various spatial factors in spatial composition to support this process.
These exploration includes object attributes such as placement \cite{palmer2008aesthetic,svobodova2014does,sample2020components,sammartino2012aesthetic}, orientation \cite{zhang2019facing,leonhardt2015your}, and relationships \cite{leyssen2012aesthetic,palmer2013visual}, as well as contextual factors such as lighting condition \cite{graham2010statistical,huang2018light,oh2024lumimood} and perspective \cite{gooch2001artistic,maqbool2023aesthetic,yang2019aesthetic}.
Human posture in spatial composition, as an essential element of 2D visual works, has been studied in visual art \cite{poore1976composition,zhang2012aesthetic,khan2012evaluating} and computer vision research \cite{cao2019openpose,cao2017realtime} as well. 

Recent advances in GenAI, while lowering the barriers for image creation \cite{brade2023promptify,chung2023promptpaint,oppenlaender2024prompting}, still lack a streamlined method to handle these factors in spatial composition. 
We propose \Oursystem{}{} to fill this gap by modeling the spatial composition in a virtual environment, collecting user intentions from it, and then interpreting the user constraints to GenAI. 
To facilitate this process, we categorized a taxonomy of spatial composition rooted in the human perception theory and established visual design research.
Unlike theoretical explorations into the aesthetic effects of individual spatial elements, this taxonomy emphasizes the practical application of common elements in image generation to faciliate user control. 
Our taxonomy consists of two distinct dimensions:

\textbf{\textit{Object Composition}} describes attributes directly associated with objects, including the transformation of individual objects and the composition of a group of objects. Specifically, the object placement, rotations, and human postures.

\textbf{\textit{Contextual Composition}} encompasses contextual factors influencing scene perception and visual effects. These include perspective, lighting direction, and illumination intensity.

We further illustrate the dimension of our taxonomy in \autoref{fig:taxonomy}.

\subsection{Controlling Spatial Composition}
Controlling spatial composition refers to handling the factors, such as object position, towards the desired outcome.
The existing controllable image generation systems adopt language and sketch to specify these factors. For example, "I want to put a car in front of a house."
While intuitive, language descriptions are typically imprecise, lacking the fine-grained control needed, thus requiring extensive trial-and-error to achieve the intended result.
On the other hand, the hand sketching method offers precision control, but requires dedicated practice to achieve such precision. 

To streamline this process and attain fine-grained spatial control, we utilize a 3D engine \cite{UnityWebsite} to model the objects and contextual factors within a virtual environment. 
The 3D engine simulates the object and renders the environment in a virtual 3D space, enabling real-time object manipulation with common interactions, such as mouse and keyboard, gesture, and voice. 
Therefore, it allows users to perceive and control the spatial composition in a native 3D space as in the real world, as shown in the \autoref{fig:taxonomy}.

However, current 3D interaction systems utilizing 3D engines are typically tailored for specific scenarios, wherein interactive objects are predefined. 
To scale 3D engines to image generation, the system must be capable of creating and managing interactive objects and environments based on the varying content of images.
This procedure involves three fundamental processes: 

\begin{itemize}
    \item \textbf{Object Registration} that creates relevant objects adherent to the user prompt.
    \item \textbf{Scene Synthesis } that initializes the scene with a reasonable spatial composition of registered objects.
    \item \textbf{Interaction Mapping} that adds interaction affordance to the registered objects and maps the user input to the object affordances.
\end{itemize}

Object registration for 3D tasks, such as modeling and animation, has been extensively explored. 
Open datasets like Objaverse \cite{deitke2023objaverse} and ShapeNet \cite{chang2015shapenet} offer diverse, generalized 3D assets. 
Users can efficiently retrieve desired objects from these datasets based on the similarity between the objects and the given contexts \cite{reimers2019sentence}. 
Additionally, recent advancements in 3D generative models provide methods for creating 3D assets directly from textual prompts \cite{dong2024coin3d,gao2024cat3d}. 
Considering both efficiency and asset quality, we opt for open 3D asset datasets to register objects.
Moreover, recent improvements in the spatial reasoning capabilities of large language models (LLMs) have inspired researchers to develop reliable scene synthesis workflows based on user prompts \cite{sun2024layoutvlm}.
By leveraging these foundational technologies, we can automate the creation of interactive virtual environments suitable for a wide range of image generation tasks.

\subsection{System Input and Intermediate Output}

After establishing a taxonomy of spatial composition and presenting our approach to controlling it, we now describe how users interact with the system (input) and how the system interprets user spatial intentions (intermediate output) to constrain GenAI.

\begin{figure}[h]
  \centering
  \includegraphics[width=\linewidth]{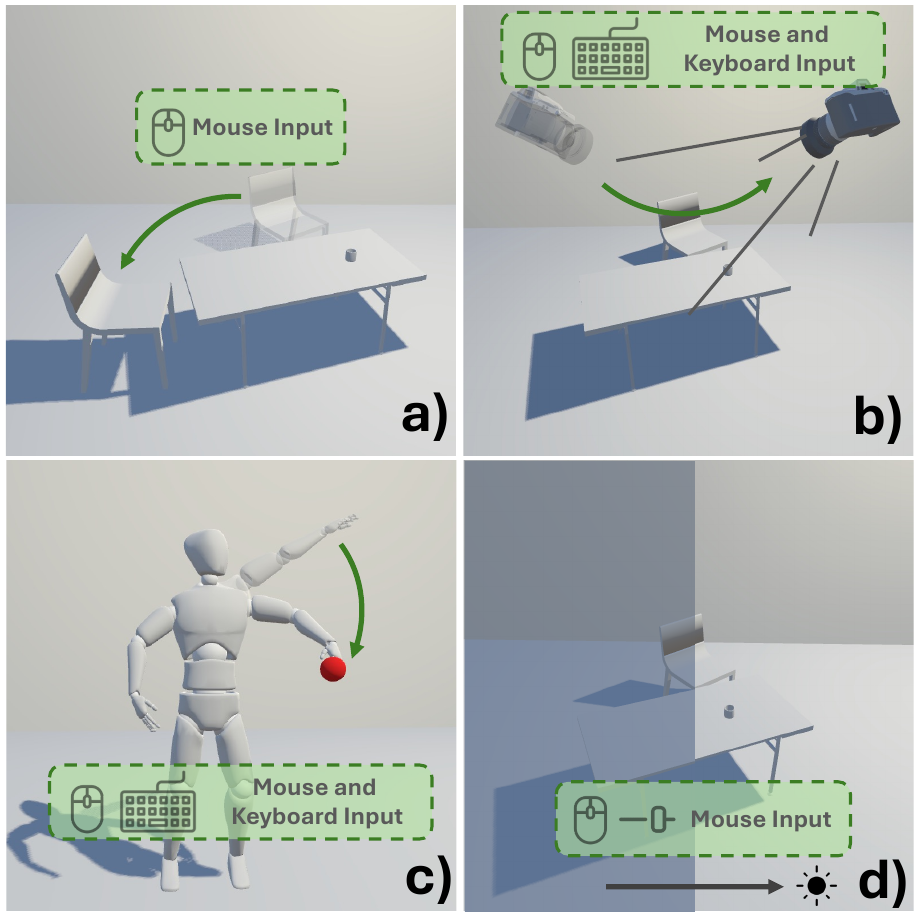}
  \caption{The mouse-and-keyboard interaction design for the system. The user can control the object using a mouse, change the perspective, or configure the human avatar with mouse and keyboard, or adjust the lighting condition by mouse and a slider.}
  \Description{None}
  \label{fig:mouse-and-keyboard input}
\end{figure}

Input methods for 3D manipulation have been extensively explored \cite{mendes2019survey,jankowski2013survey}, including traditional methods like mouse and keyboard \cite{schultheis2012comparison}, touch screens \cite{fiorella2010multi,wu2015touchsketch}, and gesture-based interactions in virtual and augmented reality \cite{alkemade2017efficiency,goh20193d}. 
Among these modalities, mouse and keyboard interactions are most prevalent within desktop environments—the primary platforms supporting image generation tasks. 
Therefore, we adopt \textbf{mouse-and-keyboard} inputs modality for their accuracy, less fatigue, and user familiarity \cite{besanccon2017mouse}.

We have developed a mapping between mouse-and-keyboard inputs and changes in the object status, as illustrated in \autoref{fig:mouse-and-keyboard input}
This mapping adopts the mouse drag and drop interaction \cite{sikkel2014clicking}  and keyboard that wildly used in 3D manipulation software \cite {claraio,substance3dstager}.  
Specifically, users can manipulate objects using mouse controls: the left mouse button translates objects, the right mouse button rotates objects, and the middle mouse button resets objects to their initial state.
Camera control is managed through mouse-and-keyboard inputs. 
Users can translate the camera horizontally using the "W, A, S, D" keys and vertically using "Q" (up) and "E" (down). 
Additionally, camera rotation is controlled using the right mouse button. 
To avoid conflicts between camera rotation and object rotation, the system detects if the cursor is interacting with an object and prioritizes object rotation in such cases.
The system provides control handles on the joints of the human avatar. Users can manipulate the avatar joints by clicking and dragging these handles, similar to object manipulation.
Illumination can be adjusted via a slider control. 
This slider appears when users hover the cursor over the light source and press and hold the right mouse button.
To implement this mapping, we developed a decision tree algorithm that maps the mouse-and-keyboard interaction affordance to the objects. 
This decision tree algorithm is further discussed in the next section.

To interpret user intentions to GenAI, the system has to output intermediate data to represent the user-defined spatial composition.
We systematically reviewed the latest conditional generative models and collected their conditioning methods.
This review informed the design of our system to \textbf{encode the conditions (spatial constraints)} that are interpretable by current generative models.
We discuss the representation of these conditions in detail in Section 4.

Therefore, our approach enables precise spatial control by (1) collecting accurate user intent through familiar mouse-and-keyboard inputs and (2) encoding these intentions into spatial constraints interpretable by generative models. 
For example, when a user positions and orients a sofa using mouse-and-keyboard controls, our system automatically generates multiple complementary outputs: a depth map representing spatial positioning and orientation, a relationship graph describing the sofa's spatial context with other scene elements, and a transformation of the camera indicating the user's perspective. These outputs then serve as direct inputs for the generative model.

In the subsequent section, we detail the technical development of our system and illustrate its workflow through a specific use-case scenario.

\section{Canvas3D}

In this section, we first walk through the system, then discuss the implementation of each module of \Oursystem{}. 
After that, we present the interface of \Oursystem{}.
Last, we discuss the hardware and software implementation of the system.

\subsection{System Overview}

The system consists of four submodules designed to implement the proposed image-generation workflow.
The process begins with a user-provided text prompt. 
Based on this prompt, the system creates an interactive 3D environment, referred to as the "\textit{canvas}", that allows users to specify their desired spatial composition.
After the user arranges objects within the canvas, the system automatically interprets this spatial composition and converts the user's intentions into input for a generative model, which then produces images accordingly. 
Users can continue editing the spatial composition on the canvas until they are satisfied with the final image.
In the following section, we discuss in detail how the system constructs the interactive canvas and interprets user spatial intentions to facilitate image generation.

\begin{figure*}[h]
  \centering
  \includegraphics[width=\linewidth]{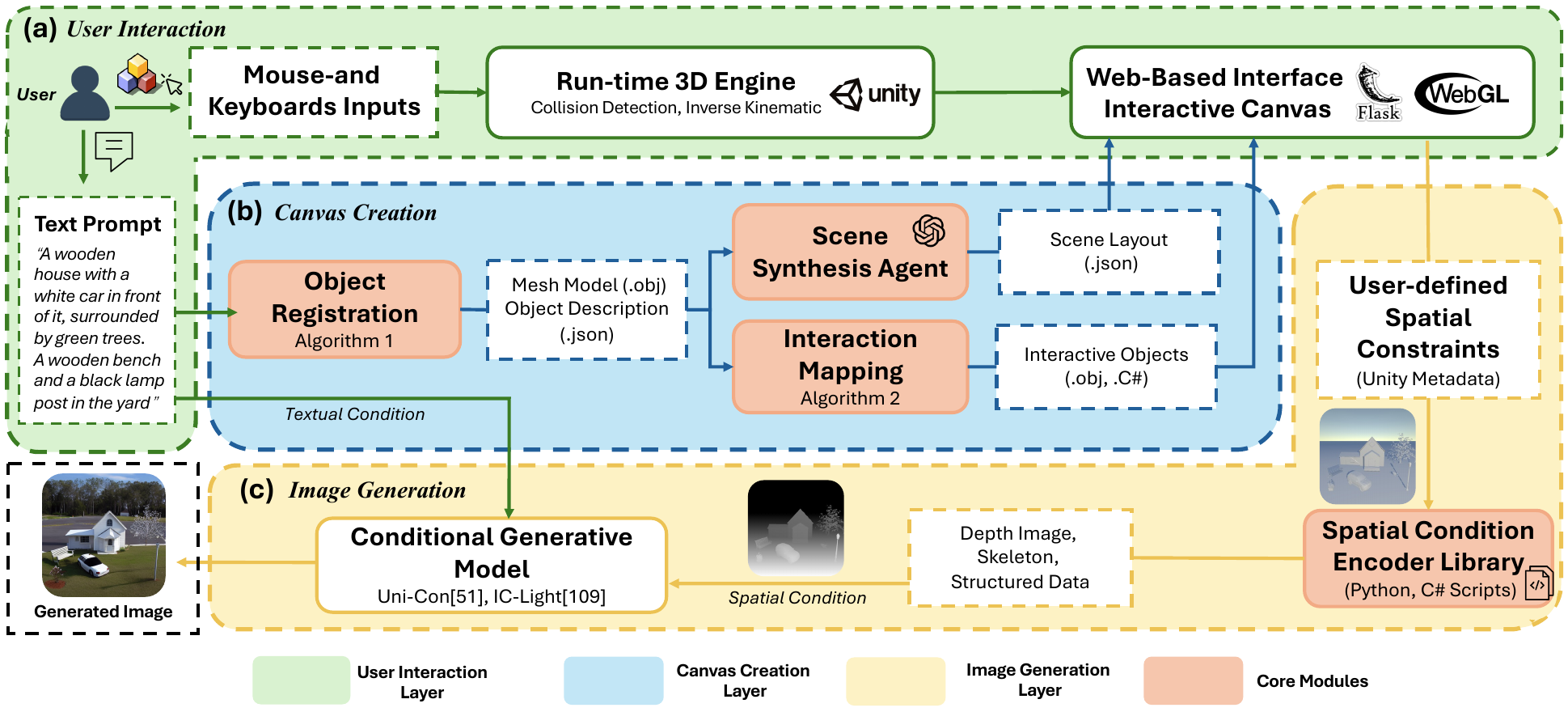}
  \caption{The system architecture of the \Oursystem{}. The system is a composition of three layers: the user interaction layer, the canvas creation layer, and the image generation layer. Each layer carries sub-modules to support the system functionality. The core modules of the canvas creation pipeline are particularly color-coded.}
  \Description{None}
  \label{systemworkflow}
\end{figure*}

\textbf{\textit{Object Registration}} 
Upon user prompt, the system first registers the relevant objects. The object registration involves gathering the object information and retrieving the corresponding mesh from a dataset. 
The collected object information, including dimension, initial rotation, scale, and native affordance, is later utilized for scene synthesis and interaction mapping.
The retrieved object meshes are used to instantiate interactive objects within the 3D environment.

\textbf{\textit{Scene Synthesis}} 
After registering the objects, the system synthesizes a scene by considering the common-sense spatial relationships among them. 
This synthesis provides users with initial object placements within the scene, which users may accept as-is or further refine by rearranging the individual objects and adjusting the overall layout.

\textbf{\textit{Interaction Mapping}} 
To support interaction, the system identifies the interaction affordances of retrieved objects and maps the user input to the affordances. 
Supported user-object interactions depend on the object's native affordances and its role within the spatial composition.

\textbf{\textit{Condition Encoding}}
Once users have arranged the spatial composition to their desired configuration, they can select a generative model to produce the corresponding image. 
The system encodes the established 3D spatial composition into constraints interpretable by the chosen generative model.

\subsection{Object Registration} 
\subsubsection{Dataset Collection}

To prototype the system, we selected 40 common indoor and outdoor object categories and curated a collection of 57,396 3D models from the ShapeNet~\cite{chang2015shapenet} and Objaverse~\cite{deitke2023objaverse} datasets, ensuring each category contains at least 300 models. 
Please refer to the supplementary material for more details.
We removed the texture and color for the models, preserving only their geometric shape information for precise collision detection within the 3D engine.
Additionally, we utilized the Cap3D model~\cite{luo2023scalable} to generate textual annotations for each 3D model. 
These annotations capture comprehensive semantic information, offering detailed descriptions that are essential for effectively selecting appropriate models.
Note that the current dataset has been collected for system prototyping purposes and can be easily expanded using the proposed workflow.

\subsubsection{Category and Quantity}
To ensure that the generated 3D scene aligns with the user's intentions, the system analyzes the user's textual input to identify the required object categories and the number of objects needed in each category. 
During this process, our system also infers implicit user intentions by analyzing the relationship between each category. 

Specifically, our system maintains a directed relationship graph to capture dependencies among categories within our dataset. 
By analyzing the input prompt and leveraging these relationships, our system can infer and generate an appropriate category list that aligns with users' needs.
For example, a relationship graph like "table → laptop" indicates that "table" is a prerequisite for "laptop." 
Therefore, if the input text mentions a laptop, it implies that a table should also be included in the generated 3D scene. 
To achieve this, we analyze and refine the list of objects using a reasoning LLM \cite{openai}. 

\subsubsection{3D Model Retrieval}
The system selects appropriate 3D models for each identified category by matching the user's input with textual descriptions of available models stored in the database. 
Additionally, the system estimates appropriate object sizes based on inferred intentions, ensuring consistency and realism in the generated scene.

To enhance consistency between 3D models and user's intended scenarios, we propose a ranking algorithm based on textual descriptions of the 3D models. 
This algorithm calculates similarity scores between each candidate model and the description of the target scene, subsequently selecting the top-ranking models. 
This algorithm effectively reduces the likelihood that objects appear in inappropriate contexts. For example, unless explicitly indicated otherwise, a clock within an indoor environment typically implies a wall-mounted clock rather than a floor-standing clock. The algorithm is shown in algorithm \autoref{alg:compute_similarity_mesh}.

\begin{algorithm}[htbp]
\caption{Iterative 3D Model Retrieval}
\label{alg:compute_similarity_mesh}
\begin{algorithmic}[1]
\Require User prompt description $D$, full categories set $C$, where each subset $C_i$ contains models $M_i$, and their annotation denoted as $M_i.anno$.
\Ensure Retrieved models $R = {r_1, r_2, \dots, r_n}$

\State $D_{curr} \gets D$
\State $R \gets \emptyset$

\For{each category $C_i \in C$}
\State $minScore \gets \infty$
\State $bestModel \gets \text{null}$

\For{each model $m_j \in M_i$}
    \State $score_j \gets \text{similarity}(m_j.anno, D_{curr})$ \Comment{Equation (1)}

    \If{$score_j < minScore$}
        \State $minScore \gets score_j$
        \State $bestModel \gets m_j$
    \EndIf
\EndFor

\State $R \gets R \cup \{bestModel\}$
\State $D_{curr} \gets D_{curr} \odot bestModel.anno$ \Comment{$\odot$ denotes textual concatenation}

\EndFor

\Return $R$
\end{algorithmic}
\end{algorithm}

The semantic similarity between two textual descriptions is computed as:

\begin{equation}
\text{sim}(t_a, t_b) = \cos\left(\mathbf{e}(t_a), \mathbf{e}(t_b)\right) = \frac{\mathbf{e}(t_a) \cdot \mathbf{e}(t_b)}{\|\mathbf{e}(t_a)\| \, \|\mathbf{e}(t_b)\|}
\end{equation}

where $\mathbf{e}(t)$ is the embedding vector obtained by encoding the textual annotation using Sentence-BERT \cite{reimers2019sentence}. The cosine similarity scores range from -1 to 1, with higher scores indicating greater semantic similarity.

\subsection{Scene Synthesis}

\begin{figure}[h]
  \centering
  \includegraphics[width=\linewidth]{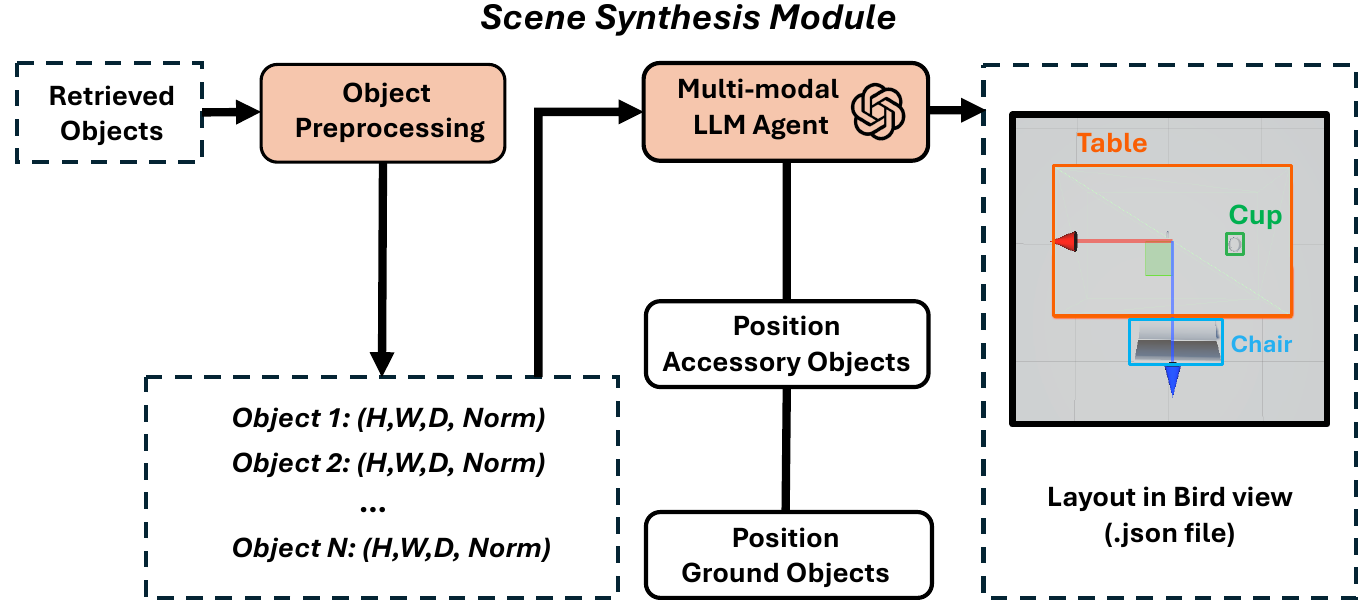}
  \caption{The architecture of the scene synthesis module. The scene synthesis module consists of two components, a object preprocessing algorithm and the multi-modal LLM agent. The preprocessing algorithm extracts the dimension of the object and the directional norm. Then, an LLM agent utilizes object information and user prompts to complete scene synthesis that provides the user a reasonable initial spatial composition in the canvas.}
  \Description{None}
  \label{Interface Integration}
\end{figure}

To minimize the user's effort in setting up the spatial composition, our system initializes the position and orientation of each model based on the given textual prompts and retrieved 3D models. 
This initialization prioritizes plausibility and coherence in the generated scenes. 
Additionally, our system ensures that the generated object placement complies with fundamental physical principles and that the overall room layout aligns with the constraints and conventions typical of common real-world scenarios.

\begin{figure*}[h!]
  \centering
  \includegraphics[width=\linewidth]{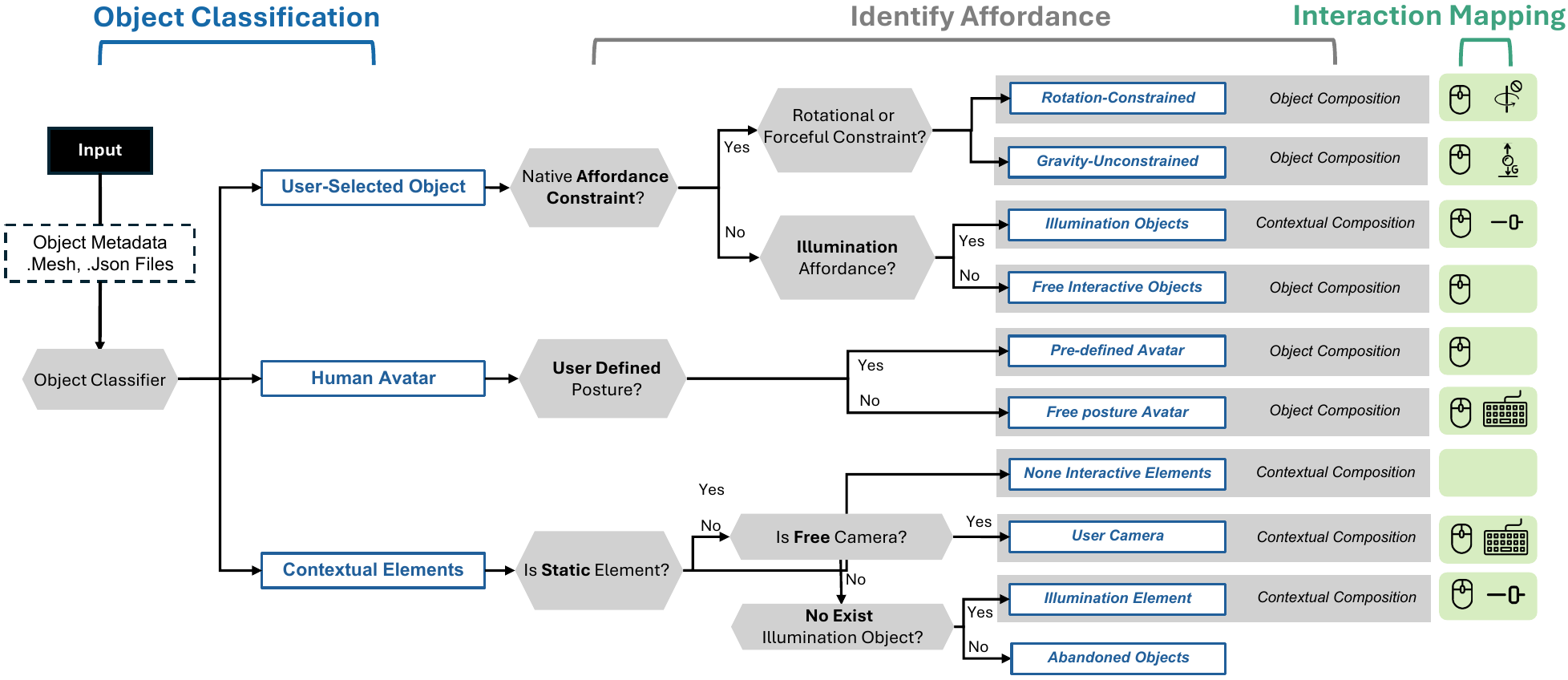}
  \caption{The decision tree algorithm (Algorithm 2) creates interactive objects by three stages: (1) classifying the input objects, (2) identifying the object interaction affordances, and (3) mapping the user input to the identified affordances. }
  \Description{None}
  \label{fig:decisiontree}
\end{figure*}

We employ a standard workflow to address this challenge by simplifying the 3D scene layout into a 2D representation from a bird's-eye view~\cite{feng2023layoutgpt, wei2023lego}.
Additionally, our system organizes objects based on their functional roles. Specifically, it first places items that stand directly on the floor(referred to as grounded objects, such as tables), then arranges objects that rest on top of these grounded objects (referred to as accessory objects, such as mugs). 

Inspired by previous work~\cite{feng2023layoutgpt}, this process leverages LLMs' reasoning capabilities and common-sense knowledge to ensure logical consistency and realism in the scene layout.

\subsection{Interaction Mapping}

To support user interaction with the 3D canvas, we employ a decision-tree algorithm to map user input to object interactions.
This approach first identifies the interaction affordances of objects based on their semantic categories and their roles within the spatial composition.
Then, based on these identified affordances, it maps user inputs onto the object affordances.

\subsubsection{Identify Interaction Affordance}
Objects carry different interaction affordances due to their distinct functional roles. 
To address this variation, we categorize objects into three distinct types:

\textbf{\textit{User-selected Objects}}: The items mentioned in the user prompt.

\textbf{\textit{Human Avatars}}: Representations of people, where interactions include posture adjustments and avatar translation and rotation.

\textbf{\textit{Contextual Elements}}: Contextual items influencing the broader scene context, typically supporting limited, environment-specific interactions (e.g., walls, floors, lighting).

\subsubsection{Interaction Mapping Decision Tree}

To effectively map user inputs to object affordances, we designed a decision-tree algorithm comprising multiple sequential decisions guided by predefined conditions.
As illustrated in \autoref{fig:decisiontree}, the decision tree operates in three primary stages: \textit{Object Classification}, \textit{Affordance Identification}, and \textit{Interaction Mapping}.
The algorithm takes as input object metadata generated during object registration and scene synthesis.
This metadata includes mesh models, descriptions, dimensions, and initial placements for each virtual object.
An example of this metadata is available in the appendix.

The decision-tree algorithm initially classifies objects into three categories using semantic labels from object registration. 
Each category further branches into finer decision layers, specifying affordances based on taxonomy associations and common-sense constraints.
For example, user-selected objects typically support movable and rotatable affordances, but specifics depend on practical constraints: a desk may rotate around the vertical axis, and a wall-mounted artwork remains fixed on the wall.
Applying these common-sense constraints simplifies user interactions, reduces unintended manipulations, and ensures greater interaction accuracy.

Finally, the system maps user input to the defined affordances.
Based on the previously discussed mouse and keyboard interactions, user inputs are translated into actions consistent with the affordances assigned to each object.
We implement this mapping according to the input-output model widely adopted for authoring object interaction affordances \cite{wang2021gesturar,zhu2022mecharspace}.
For each object, the decision tree assigns a customized script that defines its affordances and controls interactions within the 3D canvas. 
Below, we illustrate interactions determined by the decision tree.

For \textit{\textbf{user-selected objects}}, mouse controls rotation and translation as described in Section 3. Interaction constraints are applied based on fine-grained decisions; for example, rotation-constrained objects have specific rotation axes frozen, and gravity-unconstrained objects are unaffected by gravity. Illumination objects include an additional slider to control intensity. Objects labeled as "free interactive" have no interaction constraints.

For \textbf{\textit{human avatars}} with predefined postures (e.g., sit, walk), the system creates prefab avatars that users manipulate similarly to other objects. Free-posture avatars can be directly controlled with the mouse. The posture of the human avatar is controlled by an inverse kinematic algorithm \cite{tevatia2000inverse}. 

Static \textbf{\textit{contextual elements}} do not support interaction. Camera objects map keyboard inputs to interactions.
If no illumination object is detected, the system adds global illumination to the canvas. 
Uncategorized objects are excluded by the decision tree to avoid unintended elements.

\subsection{Spatial Condition Encoding}

The spatial condition encoding module translates the user's spatial intention into explicit spatial constraints interpretable by the generative models. 
To integrate with a broad range of generative models, the system collects and encodes spatial constraints into common intermediate modalities, including scene image, depth images, human skeleton postures, structured data format (JSON), and native 3D modalities (.obj files, .ply files, and scene metadata).
We implemented a function library to produce these modalities that the generative models can directly consume.
This function library is modular and can be easily extended to facilitate the integration of new generative models in the future.
The detailed descriptions of each modality are as follows:

\textit{\textit{Scene Image:}} The scene image is obtained by capturing a screenshot of the canvas. Contextual elements, such as the light indicator, are excluded from the scene images.

\textit{\textit{Depth Image:}} The system generates depth images by projecting the configured 3D scene into a depth camera view. 

\textit{Human Skeleton Data:} The system captures human skeleton data by mapping joints and the skeleton structure of a human avatar onto the OpenPose skeleton format \cite{cao2019openpose}.
Due to differences between the native avatar skeleton format and the target format, we implement an algorithm to generate missing skeleton points, such as dummy skeleton joints between the eyes and ears.

\textit{Lighting Condition Data} The system records spatial data related to lighting conditions using a structured format. Since lighting effects depend on the relative positions of viewpoint and light sources, we transform their world coordinates into a viewpoint-centered local coordinate system. The transformed coordinates and illumination intensities are encoded into a .json file compatible with a light-conditional generative model \cite{zhang2025scaling}.

\textit{Native 3D Mesh and Metadata:} The system can directly export the 3D mesh model or sample point clouds from the 3D mesh to encode spatial information into native 3D data formats.

\begin{figure}[h]
  \centering
  \includegraphics[width=\linewidth]{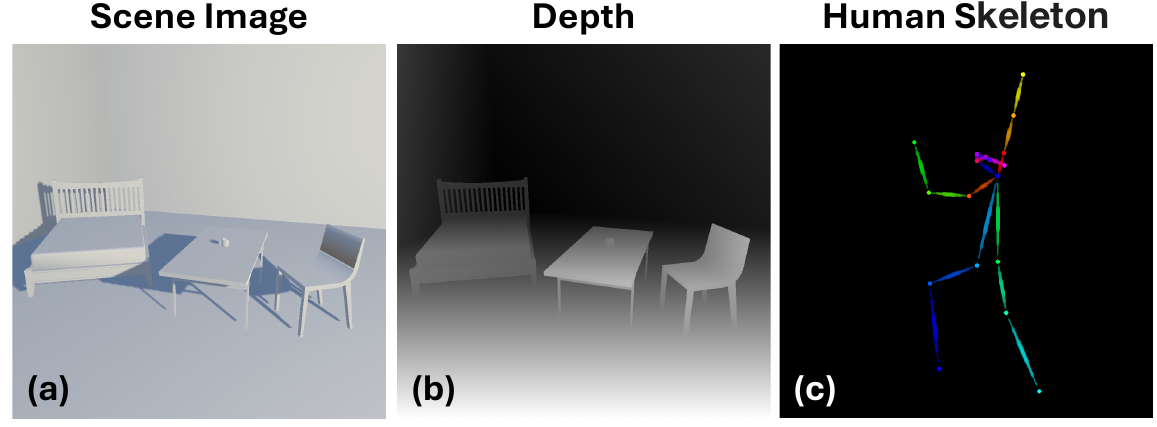}
  \caption{Example of spatial condition encoding result. The system encodes the user's spatial intent explicit conditions for the generative model, including (a) scene image (a screenshot of the canvas), (b) a depth image, and (c) a human skeleton.}
  \Description{None}
  \label{fig:spatial condition encoding}
\end{figure}

We illustrate the representation of scene image, depth, and human skeleton modality in \autoref{fig:spatial condition encoding}.

This structured encoding approach effectively captures user spatial intentions from the interactive canvas, converting them into parametric spatial constraints.
Therefore, it enables generative models to produce images accurately aligned with the user expectations.

\subsection{User Interaface}

\begin{figure*}[h]
  \centering
  \includegraphics[width=\linewidth]{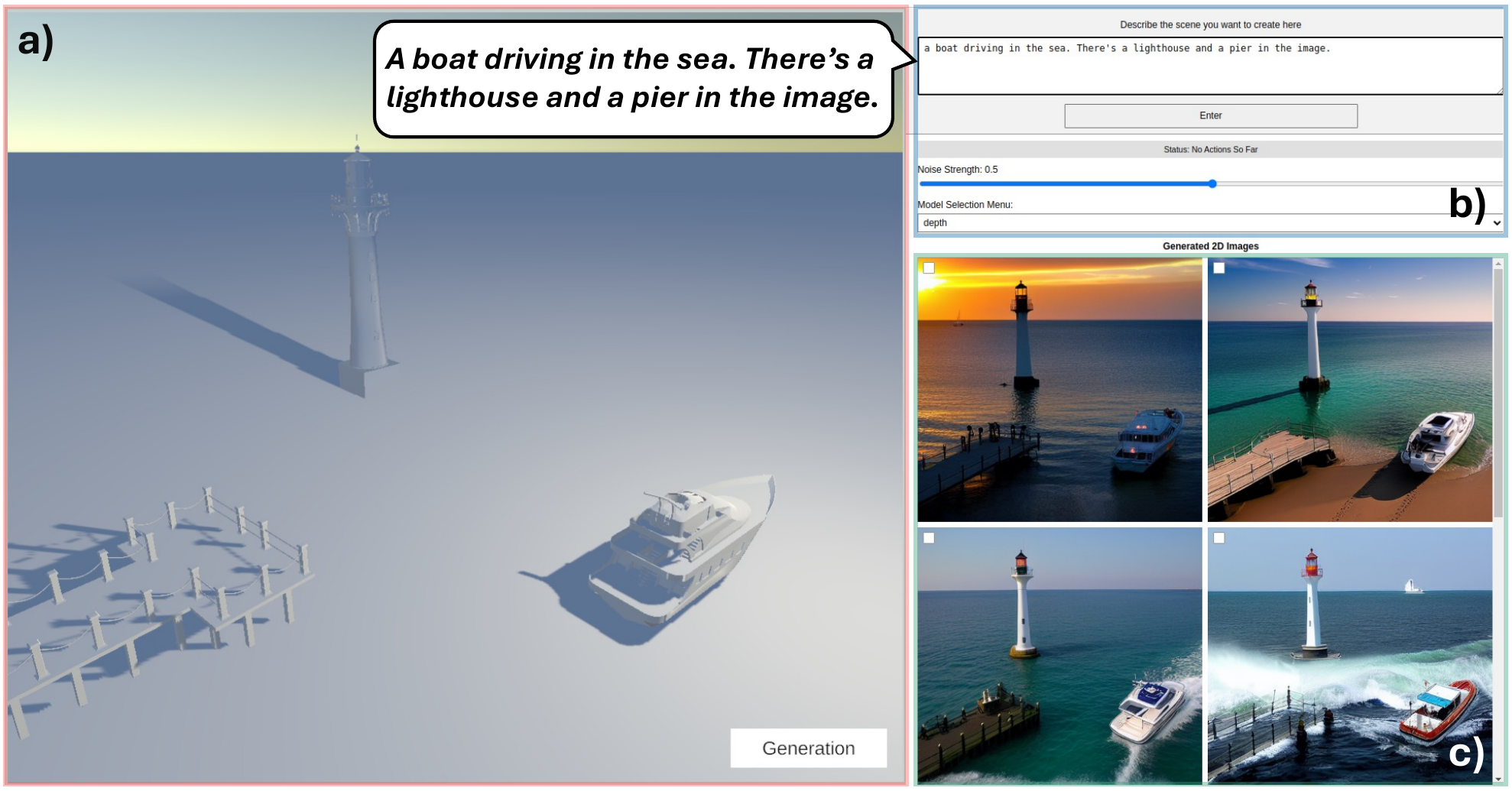}
  \caption{The user interface of \Oursystem{. The user interface consists of three components: (a) 3D interactive Canvas, (b) prompt dialog box, and (c) generated image library}}
  \Description{None}
  \label{fig:interface}
\end{figure*}

To enable intuitive and accurate image generation with the proposed workflow, we developed a web-based user interface (as demonstrated in \autoref{fig:interface}) using the Flask web framework and deployed it on a local server.

\begin{figure}[h]
  \centering
  \includegraphics[width=\linewidth]{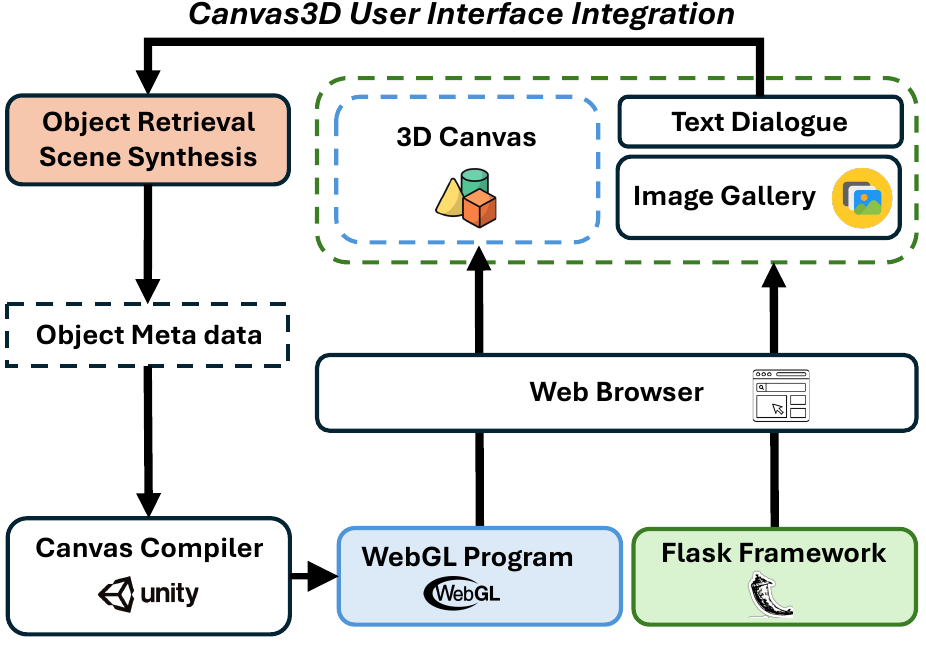}
  \caption{The architecture of the user interface integration. The \Oursystem{} integrated the 3D canvas with a front-end interface developed. The 3D canvas is driven by a WebGL program compiled by the Unity 3D engine. It is integrated with a front-end interface developed on the Flask framework. A web browser hosts the two components to facilitate user interaction.}
  \Description{None}
  \label{fig:Interface Integration}
\end{figure}

To integrate the interactive 3D engine-driven canvas, we compiled our visual canvas developed in Unity into a WebGL program, as demonstrated in \autoref{fig:Interface Integration}
This approach allows the user to interact with the object within the interactive canvas and other interactive web elements necessary for the image generation.

\subsubsection{Interface Components}

The interface comprises three primary components designed to assist users through the image generation process. 
(1) The \textit{Prompt Dialog Box} serves as the entry point for users to initialize the scene. 
Users can input textual prompts to generate an initial spatial composition within the interactive canvas and use this dialog to guide subsequent image generation further. 
(2) Central to user interaction, \textit{3D Interactive Canvas} provides a real-time, 3D engine-driven environment. 
(3) \textit{Generated Image Library} displays generated images, allowing users to easily browse results, compare variations, and mark preferred outcomes. 

\subsubsection{Supplementary Functions}

To further streamline interaction and enable flexibility, the interface encompasses additional supportive features.
The User can dynamically add new objects via textual prompts or upload a customized mesh model to the canvas. 
To increase diversity in generated images, the user may also inject controlled randomization into objects to explore a wider design space on the aesthetic features. 
 A model selection button allows users to easily switch between available generative models. 

\subsection{System Implementation}

The Canvas3D system is implemented as a web-based application utilizing Unity WebGL for graphical rendering, a Flask-based front-end user interface, and a back-end server comprising Unity (version 6000.03f1) and Python (version 3.8.20). 
The system adopts GPT-o1 as the LLM agent for scene synthesis.
The whole system was run on a PC (Intel Core i9-10900KF CPU, 3.7GHz, 128 GB of RAM, and an NVIDIA GeForce RTX 3090).

\section{User Study Evaluation}

\Oursystem{}{} aims to tackle the problem of precise spatial control in generated images. 
We conducted a two-session user study to evaluate system performance based on the considerations outlined in previous sections. 
Specifically, we examined: 

\begin{itemize}
    \item \textit{How well the system supports users in configuring image objects and contextual elements.}
    \item \textit{How accurately the generated images align with user-specified spatial constraints using our system.}
    \item \textit{The quality of the final images and overall system usability.}
\end{itemize}

In the first session, we conducted a closed-ended experiment to compare our system against a baseline. 
This comparison aimed to evaluate the spatial control, result quality, and user experience of two systems.
The second session was an open-ended experiment in which the participant could use our "in the wild" system to generate images.
This evaluation aimed to assess the functionality of the system modules and the overall usability.

We recruited 12 participants (Male=7, Female=5) from a university community. 
None of the participants had prior experience with our specific system. 
Two-thirds (Number=8) have used generative models for image creation. 
To acquire insights from experts, visual design experts (Number=5) from visual art, design, and computer graphics majors were involved.
The entire user study lasted approximately 1.5 hours, and each participant was compensated with a \$30 e-gift card for their participation.
Prior to the main tasks, each participant received introductory guidance to familiarize themselves with the interface and functionality of the system. 
During the actual study tasks, no additional guidance or intervention was provided; participants independently followed the requirements to complete the task. 
Upon completing both sessions, participants were asked to fill out a 7-point Likert-scale questionnaire evaluating their experiences, along with a standard System Usability Scale (SUS) questionnaire and a NASA Task Load Index (NASA-TLX) questionnaire \cite{hart1988development} to measure usability and perceived workload. 
Finally, we conducted conversational interviews with each participant to gather qualitative, subjective feedback about their experience with the system.
The user study was conducted under a university-approved IRB. 

\subsection{Session One: Close-Ended Experiment}

In this session, we aim to evaluate the spatial control using our system compared to a baseline system. 
We conducted a closed-ended experiment on the two systems to assess the spatial accuracy, task fluency, and quality of the generated images.
Specifically, we provide participants with a target image, as shown in \autoref{fig:Masked Target Image} that encompasses different spatial compositions. 
We ask them to generate a new image using the system with the same spatial compositions as the given one.
Additionally, the participants have different experiences in using generative models or practicing image generation tasks. 
To reduce the individual differences in the experiment result, we design the experiment as a with-in-subject study where each participant experiences all experimental conditions: using our system and the baseline system.

\subsubsection{Baseline}

\Oursystem{} aims to facilitate precise spatial control for image generation.
Therefore, our baseline system is designed based on the existing approaches for controlling the spatial factors of images.
The baseline system is equipped with a sliders feature, as discussed in recent works \cite{eldesokey2024build,bhat2024loosecontrol}, where users operate sliders to control the object positions and orientation to constrain the generative model, as shown in \autoref{fig:baseline interface}.
We adopted the same conditional generative model \cite{li2024simple} as a generation backbone for the baseline and our system.
The backbone model supports the spatial conditions created by both systems.

\subsubsection{Procedure}

The within-subject study requires the participants to experience all the conditions.
Such conditions may cause the learning effect that users perform better on the later experiment conditions. 
To mitigate the learning effect, we experiment in a counterbalancing order, where the order of conditional groups is randomized for each participant. 
During the experiment, each participant is provided a target image, where the key objects are marked to indicate the target spatial composition. 
To complete the task, each participant must generate an image with a spatial composition as close as that in the given target. 
Users can generate as many times as they prefer to acquire the final result. 
However, we set a time limitation of fifteen minutes to ensure consistency across participants. 
Following the previous procedure, each participant has two sub-tasks of image generation, and we offer them a five-minute break time between each subtask to prevent fatigue.
Once the two sub-tasks are completed, each participant will complete a questionnaire to collect subjective quantitative data.  
After the first session of our user study, the participants have a ten-minute break before starting the second session.

\subsubsection{Metrices}

In the experiment, participants generated images using both our system and the baseline system.
To quantitatively evaluate the system's ability to capture spatial relationships and leverage human intentions, we selected five metrics.
For all metrics, higher values indicate better performance. 
The quantitative data was obtained by comparing the target images, the input text prompt, and the corresponding generated images. 
Inspired by previous research \cite{huang2025t2i}, we adopted four metrics to evaluate the quality of the generated images:

\textit{\textbf{CLIP Score}}: the overall semantic similarity between the user's prompt and the generated image.

\textit{\textbf{GPT Spatial Score}}: the spatial relationships described in the user's prompt are represented in the generated image.

\textit{\textbf{GPT-CLIP Score}}: the similarity between the generated images and the target images calculated using CLIP scores between their ChatGPT-generated captions.

\textbf{\textit{Uni-Det Score}}: the consistency of spatial relationships between the target images and the generated images. Please refer to the supplement for more details.

Furthermore, we introduced an additional quality metric to evaluate the quantity aspect of the generated images:

\textit{\textbf{Recall Score}}: the proportion of categories successfully generated to the total number of intended categories.

To evaluate the task fluency, we designed a questionnaire based on the NASA TLX. 
The questionnaire collects the participants' feedback on the system features and the workload using two systems.
For more details about quality metric calculations, please refer to the appendix.

\subsection{Session Two: Open-Ended Experiment}

The second user study aimed to assess the functionality of system modules and the overall system usability. 
The evaluations were based on the quantitative and qualitative feedback collected from the questionnaire and post-study interviews. 
To acquire such feedback, we conducted an open-ended experiment that allowed the participant to use our system in the wild.

\subsubsection{Procedure}

In the open-ended experiments, participants were asked to freely generate images using our system "in the wild".
To encourage participants to engage with all system features, we introduced two loosely constrained subtasks:

\textit{(1) Generate an image involving humans to explore posture, lighting, and camera control.}

\textit{(2) Generate an image depicting a scene to explore object and perspective control.}

Participants had the freedom to choose the topics and content of their generated images, with the only restriction being the exclusion of violent and abusive materials.
We provided an optional five-minute break between the two subtasks.
The user study ended up with a questionnaire and conversational-based interview.

\section{User Study Results}

\subsection{Result of Close-Ended Experiment}
In this section, we present the results of the closed-ended evaluation. The objective evaluation data was collected using a Likert-type questionnaire, while the subjective evaluation was derived from an analysis of task procedures and image generation results. For the two systems evaluated, we report the mean scores (M), standard deviations (SD), and significant p-values (p) for comparison.
We performed the Shapiro–Wilk test on each group of data samples to assess normality and found that most paired data significantly deviated from normal distributions. 
Consequently, we utilized the Wilcoxon signed-rank test, a widely adopted non-parametric method for paired data, to evaluate whether there are significant differences between those two systems and reported the corresponding p-values.

\begin{figure*}[h!]
  \centering
  \includegraphics[width=1\linewidth]{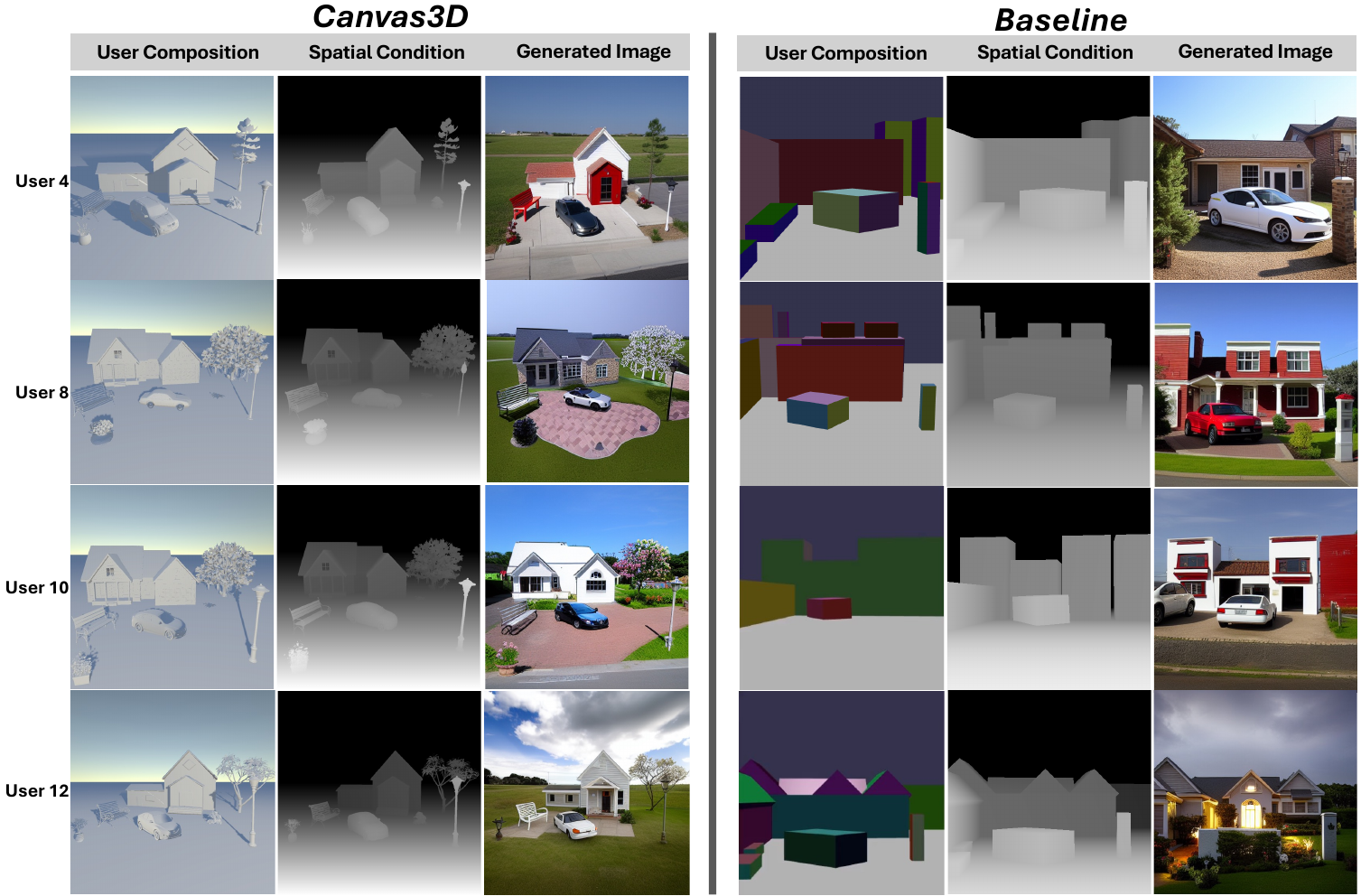}
  \caption{The example of comparing user-generated images using \Oursystem{} (left three columns) and a baseline (right three columns) in the close-ended experiment. For each system, the screenshot of user composition on the interface, the spatial condition for the generative model, and the final generated image are presented. }
  \Description{None}
  \label{fig:user study closed-ended}
\end{figure*}

\subsubsection{System Feature Comparision}
The Likert-type questionnaire rating compares the \Oursystem{} and the baseline, as shown in \autoref{fig:close-ended questionnaire}.
In general, the participant rated our system significantly higher in terms of interactivity (\Oursystem{}: M=5.83, SD=0.83; Baseline: M=4.00, SD=2.09; p=0.030). 
\textit{"I feel like drag and drop is a lot easier because I feel like that's what we do in real life. We pick something up, move it; it just feels the same with the mouse (P5). " }

We observe that users using the baseline often conduct a trial-and-error process to find the right slider to control the degree of freedom they want. 
\textit{"I often forgot which slider controlled the depth, so I had to play around with it to see if I was using the right one (P3)."}
Similarly, our system demonstrated significantly improved spatial control on the final image (\Oursystem: M=6.25, SD=0.97; Baseline: M=4.83, SD=1.47; p=0.020).
Furthermore,  the users notably prefer interaction with actual object models instead of bounding boxes, as depicted in the object quality score (\Oursystem: M=6.25, SD=0.62; Baseline: M=4.58, SD=1.73; p=0.008). 
As one user stated, \textit{"Being able to see the actual object instead of just the bounding box is really helpful, because it lets me get the exact shape and size right when I'm placing it with other objects. It is easier for me to envision the overall composition.(P12) "}

\subsubsection{Workload Comparison}
We compare the workload to complete the tasks using the NASA-TLX liker-type questionnaire, as shown in \autoref{fig:close-ended questionnaire}.
While no statistically significant differences emerged in terms of physical demand (\Oursystem: M=2.08, SD=1.51; Baseline: M=3.33, SD=2.31; p=0.156) and mental demand (\Oursystem: M=2.75, SD=1.66; Baseline: M=3.67, SD=1.87; p=0.100), both showed a trend toward reduced workload with our system.
However, significant differences emerged favoring our system regarding perceived performance (\Oursystem: M=5.00, SD=1.04; Baseline: M=2.83, SD=1.47; p=0.005), effort (\Oursystem: M=2.42, SD=1.16; Baseline: M=4.67, SD=1.67; p=0.002) ,and frustration (\Oursystem: M=2.17, SD=1.19; Baseline: M=3.75, SD=2.14; p=0.016). 
Like one user stated, "I'm more of a visual-based person, so it was a little bit harder to work based on, like, the numbers and the sliders (P9)." \textit{And another echoed, "You had to use sliders, which was pretty cumbersome. But with a free camera, you can just move around the scene, and it makes placing objects exactly where you want them so much easier (P10)."} 
These results clearly indicate that the proposed system significantly reduced the perceived efforts and frustration while enhancing the overall performance on the image generation.

\subsubsection{Spatial Control Comparison}

We quantitatively evaluate the spatial controllability of the two systems by comparing their generated images against the target image.
Both systems utilize the same backbone architecture for image generation.
Therefore, comparing the generated results, especially their spatial compositions relative to the target images, reveals how well each system supports user spatial control.
\autoref{tab:evaluation on generated images} presents the results of this comparison. 
\Oursystem{} consistently outperforms the baseline across all the metrics.
The result shows a significant difference in the GPT-CLIP (p=0.0024), Uni-Det (p=0.0034), and Recall (p=0.0005).

The better image generation result using our system can be attributed to several factors. 
Firstly, we observe that the baseline system shows lower performance due to user frustration and interface difficulties. 
Some users complete the baseline system task with some objects missing in the image, as shown in the \autoref{fig:user study closed-ended}.
This negatively impacts both Recall and the GPT-CLIP score.

Secondly, the bounding box representation employed in the baseline system lacks sufficient accuracy to effectively communicate the user's spatial intentions. 
The visual similarity between bounding boxes makes it challenging for the generative model to clearly differentiate objects and correctly associate them with user prompts, resulting in significantly lower performance scores (Uni-Det, p = 0.0034). 
This issue becomes particularly significant in complex scenes, where closely positioned bounding boxes are frequently misinterpreted as a single object—for instance, multiple objects in the yard being mistakenly merged into a single house, as depicted in the \autoref{fig:user study closed-ended}.
Overall, these findings suggest that Canvas3D provides better spatial controllability compared to the baseline, especially evident in metrics sensitive to spatial composition accuracy.

\begin{table}[htbp]
\centering
\caption{Result of Quantitative Evaluation on Generated Images. The symbol $\uparrow$ indicates that a higher value of the metric is better}
\label{tab:evaluation on generated images}
\begin{tabular}{l |cc |cc |cc} 
  \toprule
  & \multicolumn{2}{c|}{\textbf{Canvas3D}} 
  & \multicolumn{2}{c|}{\textbf{Baseline}} 
  & \multicolumn{2}{c}{\textbf{Statistics}} \\
  & \textbf{Mean} & \textbf{Std} 
  & \textbf{Mean} & \textbf{Std} 
  & \textbf{p}    & \textbf{Sig} \\
  \midrule
  CLIP ($\uparrow$)       & 30.07 & 2.70 & 29.17 & 3.16 & 0.3911 & \\
  GPT Spatial ($\uparrow$) & 66.67 & 11.86 & 55.00 & 18.48 & 0.1855 & \\
  GPT-CLIP ($\uparrow$)    & 75.75 & 3.58 & 64.44 & 6.63 & 0.0024 & \multicolumn{1}{c}{*}\\
  Uni-Det ($\uparrow$)     & 74.37 & 12.40 &  35.80 & 27.53 & 0.0034 & \multicolumn{1}{c}{*}\\
  Recall ($\uparrow$)      & 86.57 & 10.75 & 42.83 &  10.67 & 0.0005 & \multicolumn{1}{c}{*}\\
  \bottomrule
\end{tabular}
\end{table}

\subsection{Result of Open-Ended Experiment}
In this section, we present the results of the open-ended evaluation. The evaluation was conducted using two questionnaires: the System Feature Questionnaire and System Usability Questionnaire. For each questionnaire, we analyze the mean scores (M) and standard deviations (SD) to summarize user responses.

\begin{figure*}[htp]
  \centering
  \includegraphics[width=1\linewidth]{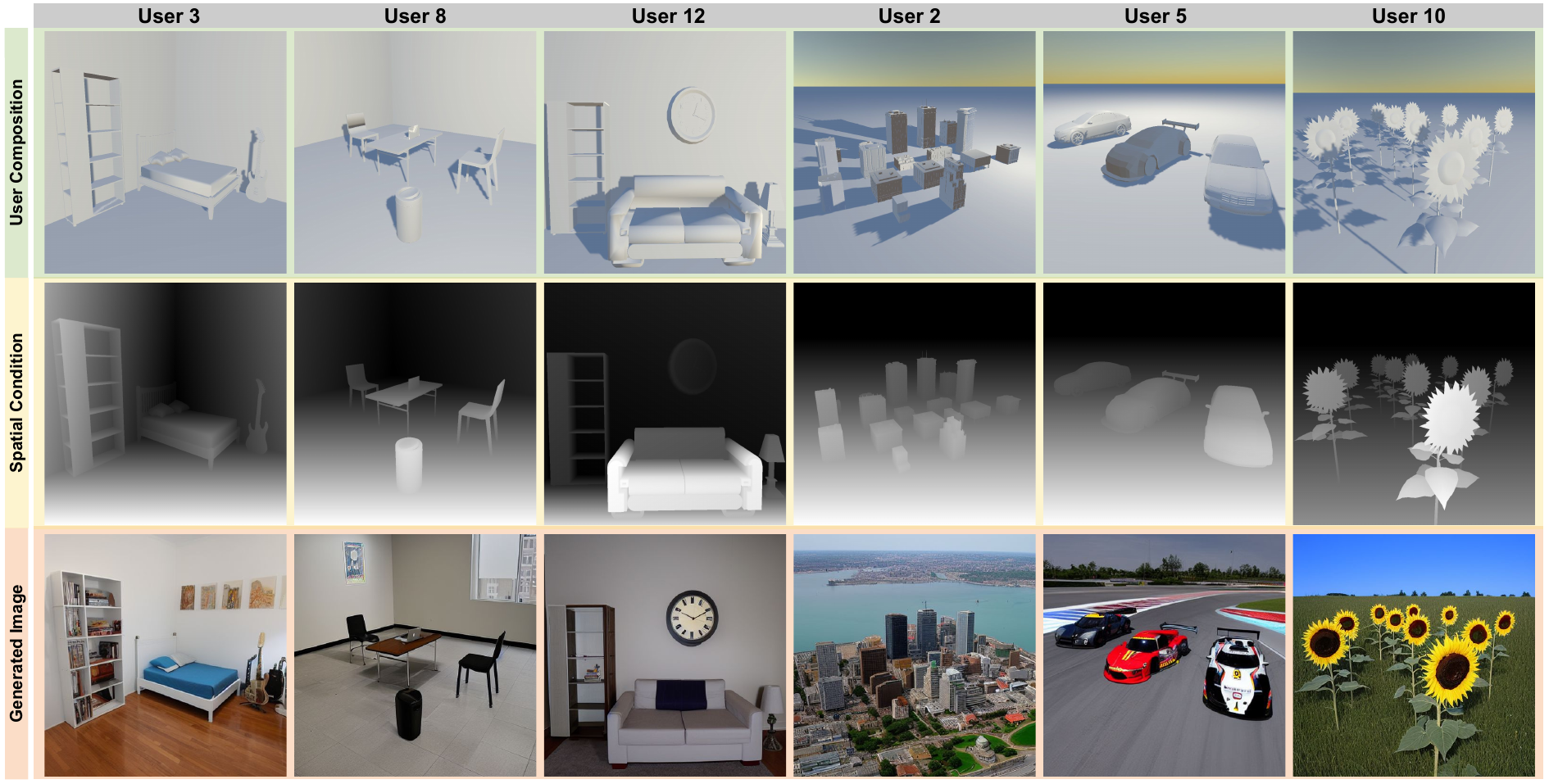}
  \caption{The example of user-generated images using \Oursystem{} for scene depiction in the open-ended experiment. }
  \Description{None}
  \label{fig:user study open-ended scene}
\end{figure*}

\subsubsection{System Usability Questionnaire}

Participants reported overall positive experiences with the evaluated system, providing favorable ratings across multiple aspects, as depicted in \autoref{fig:questionnaire of open-ended experiment}. 
The report highlight the eas of use for setting up lighting conditions, receiving the highest among all the aspect evaluated  (M = 6.42, SD = 0.76).
The lighting condition control is especially appreciated by the expert designers over the novices, as one commercial art expert mentioned, \textit{"The most important element is lighting. Initially, we think relatively less about the details, 'where should I place a computer, or how it looks like'. Instead, we're more focused on capturing the general feeling, the lighting, the visual, and the overall effect (P12)."}. Another game designer echoed, \textit{"Adding the lighting effect takes times and money. And this (control the light in reference images) will definitely save a lot of that from unecessary hours on the project (P9)."}
They found it straightforward to set up the human pose using the provided controller (M = 5.25, SD = 1.30). 
The participants appreciated the system’s capability to accurately retrieve relevant objects based on their text prompts (M = 5.25, SD = 1.23) and found object manipulation intuitive using the provided interactions (M = 5.42, SD = 1.32).  \textit{"I have no problem with the objects it gives me—I actually like the feature where it suggests more things, and I can add or remove stuff later. My ideas might change as I'm creating the scene (P3)."}
The initial image layouts generated by the system were viewed as reasonable and effective (M = 5.50, SD = 1.71). The participant echoed our adhere to the low-friction design principle, \textit{"Actually the beginning layout it gives is pretty good, it save a lot of time to organize from scratch and you can still customize the detail, like putting my laptop on the bed (P2)." }
Additionally, the simulation applied to objects was rated highly for accuracy (M = 5.50, SD = 1.32) and was considered particularly helpful for setting up spatial compositions (M = 5.58, SD = 1.38).
The interface effectively facilitated clear arrangement of the overall spatial composition (M = 5.33, SD = 1.75), and participants felt that their spatial intentions were accurately represented in the generated images (M = 5.33, SD = 1.25). \textit{"I think the image generation was very good and I think everything was fairly easy to use. (P11)."}
Participants expressed satisfaction with the quality of the generated images (M = 5.33, SD = 1.18)

\begin{figure*}[htp]
  \centering
  \includegraphics[width=1\linewidth]{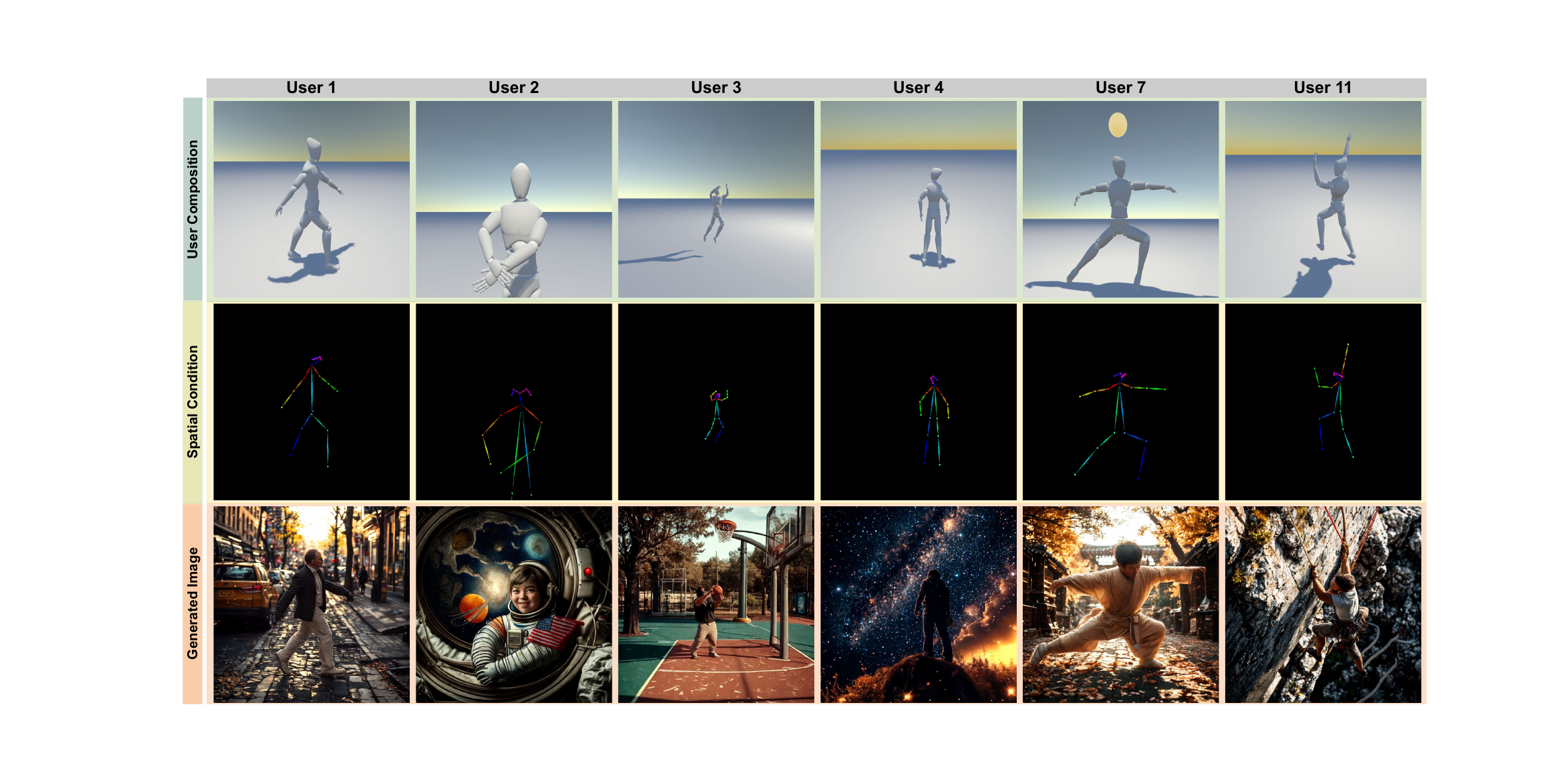}
  \caption{The example of user-generated images using \Oursystem{} for pose depiction in the open-ended experiment. }
  \Description{None}
  \label{fig:user study open-ended pose}
\end{figure*}

\begin{figure*}[htp]
  \centering
  \includegraphics[width=0.9\linewidth]{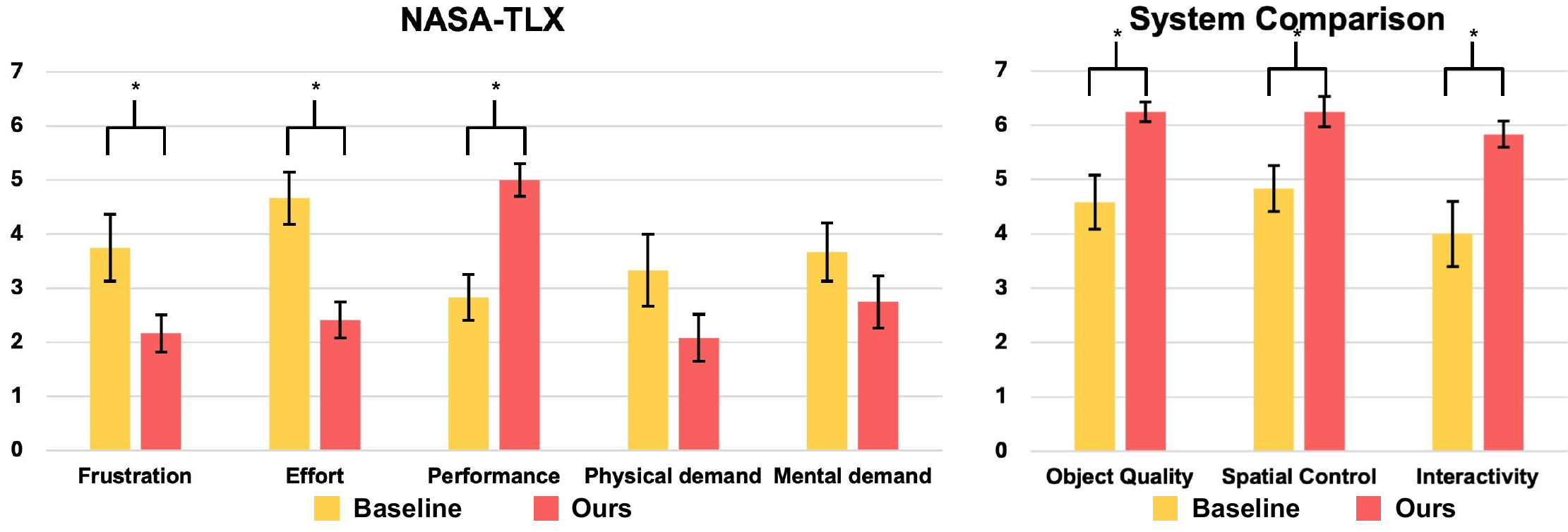}
  \caption{NASA-TLX and system comparison. Our system demonstrated significantly improved spatial control (\Oursystem: M=6.25, SD=0.97; Baseline: M=4.83, SD=1.47; p=0.020), interactivity (\Oursystem{}: M=5.83, SD=0.83; Baseline: M=4.00, SD=2.09; p=0.030), and the object quality (represented as object models in our system and as bounding boxes in the baseline system) (\Oursystem: M=6.25, SD=0.62; Baseline: M=4.58, SD=1.73; p=0.008) on the final image.
  Our system also demonstrates significant improvements on the self-perceived performance (\Oursystem: M=5.00, SD=1.04; Baseline: M=2.83, SD=1.47; p=0.005), and significantly reduces the frustration (\Oursystem: M=2.17, SD=1.19; Baseline: M=3.75, SD=2.14; p=0.016) and effort  (\Oursystem: M=2.42, SD=1.16; Baseline: M=4.67, SD=1.67; p=0.002) to complete the task.
  }
  \Description{None}
  \label{fig:close-ended questionnaire}
\end{figure*}

\begin{figure}[htp]
    \centering
    \includegraphics[width=0.5\textwidth]{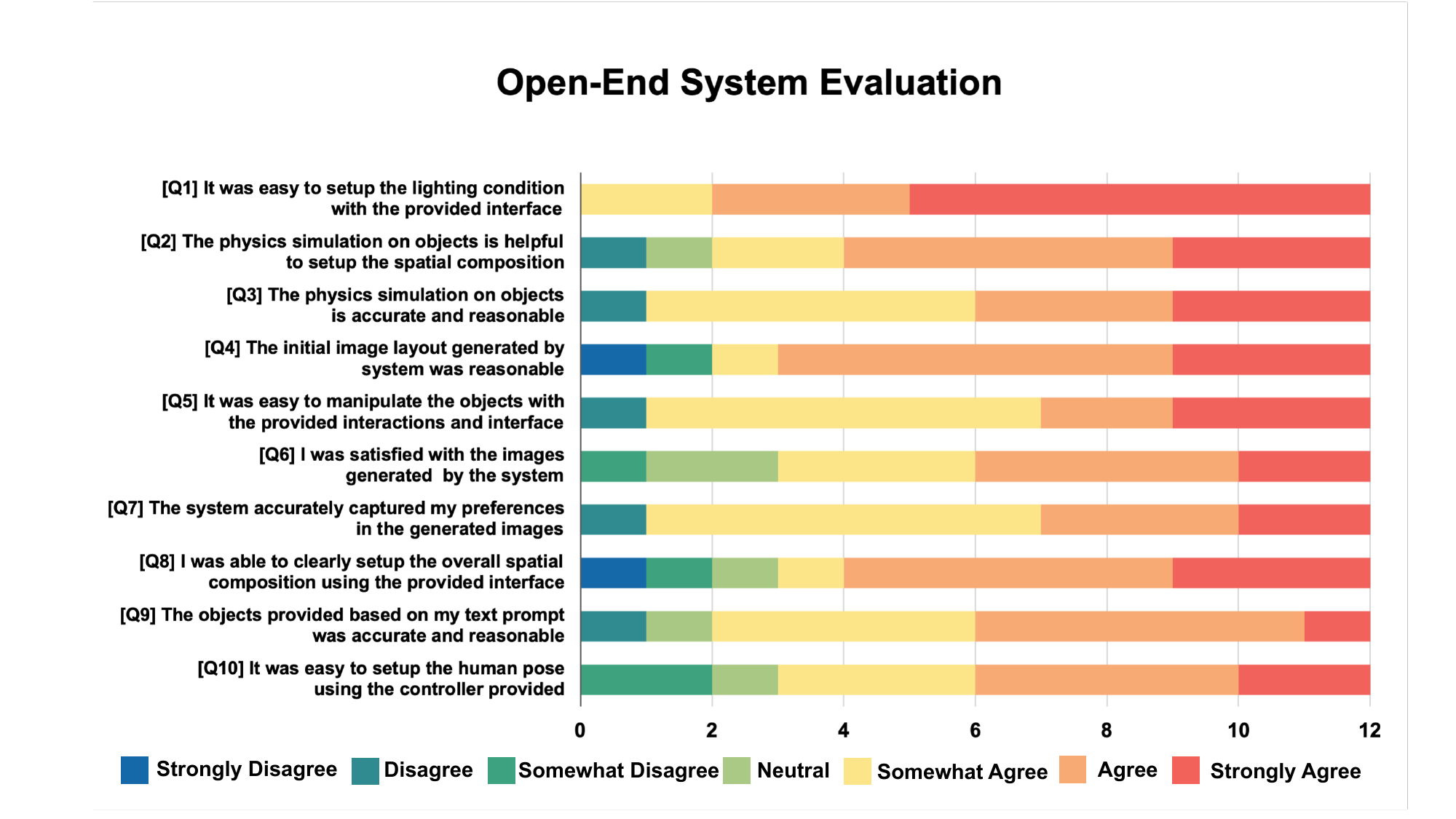}
    \caption{The result for the Open-Ended Experiment: The Likert-type questionnaire on system evaluation}
      \Description{None}
    \label{fig:questionnaire of open-ended experiment}
\end{figure}

\subsubsection{System Usability Questionnaire}

The usability evaluation indicates that our system has strong overall usability, achieving an overall usability score of 82.22. Detailed results from the system usability questionnaire are provided in the appendix.

\subsubsection{Example of User-Generated Images}

We present the generated image of six users during the open-ended experiment. 
The examples represent the system performance on: (1) generating an image of a human, as shown in \autoref{fig:user study open-ended scene}, (2) generating an image to depict a scene, as shown in \autoref{fig:user study open-ended scene}.
\section{Discussion, Limitation and Future Work}

\subsection{Improving the 3D Canvas for Comprehensive Spatial Composition}

In essence, our interactive 3D canvas supports users to precisely control the spatial composition categorized in our taxonomy.
From the user study, the participants found it easy to manipulate the spatial composition and express satisfaction with the objects provided in the canvas.
We believe some aspect of the interactive 3D canvas could be further improved to support a more comprehensive spatial control. 

\textbf{\textit{User customized constraints and affordances. }}
Now, the decision tree algorithm assigns fixed constraints and interactions to the object. 
For example, wall-mounted objects are exempt from gravity, so the user can place them in mid-air.
This design aligns the system-provided object interactions with common-sense object affordances, and it is appreciated by all users. 
We believe that allowing users to define object constraints and customize affordances would increase flexibility and facilitate creativity using generative models \cite{mi2025think,duan2024conceptvis,choi2024creativeconnect}. 
\textbf{\textit{3D Model Generation for Interactive Canvas: }}The system currently uses object retrieval and allows the user to upload mesh files to construct the interactive 3D canvas. 
This design intends to let the user focus primarily on spatial composition rather than waiting for object creation and canvas generation while preserving the flexibility to use customized objects. 
We think this feature can be improved by leveraging the advances in 3D model generation techniques\cite{zhao2025hunyuan3d, zeng2024paint3d}. 
These emerging techniques offer the generation of high-quality and articulated objects, which can potentially help the system to create a more interactive environment for users to constrain the generative model. 
With these improvements, our system could evolve into a comprehensive spatial manipulation tool, directly fulfilling users' spatial composition needs without reliance on a separate image generation model to produce the desired 2D images. 

\subsection{Expanding Spatial Control Across Interaction Modalities}

As hypothesized, the direct spatial control using mouse and keyboard was favored by all users and produced better results in the generated images, compared to the baseline. As described by one user: 

\begin{quote}
\textit{"I feel like drag and drop is a lot easier because I feel like that's what we do in real life. We pick something up, move it; it just feels the same with the mouse (P5). " }
\end{quote}
This perspective highlights how the approach aligns with the natural ways in which humans manipulate objects and offers a flat learning curve for users without any experience in 3D manipulation. 

We believe the system can be further improved by incorporating additional input modalities, such as touchscreen \cite{reisman2009screen} and gesture control \cite{kim2016touch}, to accommodate a broader spectrum of users.
The existing web-based framework enables flexible integration with other terminals by deploying the 3D engine  and generative model on a cloud server. 
However, additional experiments are required to thoroughly evaluate the interaction design across different devices and modalities.
Additionally, immersive environments using AR/VR devices provide more natural and intuitive spatial control \cite{liu2023instrumentar, duan2025parametric}.
Such environments leverage native spatial interactions based on daily gestures that humans are already familiar with, such as picking up and moving an object using two fingers.
Investigating the immersive environment method could potentially reduce the cognitive load and learning curve observed in our desktop implementation.

\subsection{Accurate Translating User Spatial Intentions}

During the user study evaluation, we found that the users are satisfied with the spatial intention captured by our system. 
We adopted the spatially accurate intermediates, such as a depth image \cite{cai2024spatialbot}, to translate the user's spatial intention to generative models.
However, as one user pointed out, depth images have limitations in some cases, for example, if the two objects are too close to each other in the depth direction, then they will have similar values in the depth image, making them difficult to distinguish by the generative model. 
\begin{quote}
    \textit{" Sometimes the system won't be able to recognize the car if I put it close to the house (P11)."}
\end{quote}
Moreover, the system processes the user's spatial intentions using separate generative models—for example, one model for posture generation and another downstream model to control lighting conditions. 
This distributed approach increases processing time and introduces inconsistencies in style and object representation within the generated images.

We envision that these limitations can be addressed by advances in conditional generative models capable of handling multiple spatial inputs, as demonstrated by recent methods\cite{wang2024instancediffusion, zhang2023controllable, mo2024freecontrol}.

\subsection{Additional Features for Expert Designers}

The current system aims to streamline the image generation process and maintain precise spatial control.
It adopts the most common spatial composition to cover the need for general image generation. 
Our system is designed not as a competitor to the current productivity tools but as a complement to lower the barrier of image creation. 
All users have appreciated the ease of use of the system features. 
However, some additional features could further benefit the experts integrating our system into professional design workflows. 
As one visual designer expert pointed out:

\begin{quote}
\textit{”It would help to have handles on the object for visually dragging or rotating along specific axes—it’s more intuitive. Also, numeric inputs next to an axis indicator are useful for precise movements. For example, if I place a cup and just want to move it slightly backward, typing a number is easier than dragging it again (P9).” }
\end{quote}

Additionally, in contrast to the everyday user, who appreciates the easy manipulation of the object, the expert emphasizes system features that can enhance perception of the overall layout, such as the free camera and the illumination control.
One expert specifically emphasized the importance of lighting control within professional workflows, stating:

\begin{quote}
\textit{"This is the key question in creative workflow: what should come first, what should come later? For me, lighting and shadows come first, followed by the placement of objects, and then finer details afterward (P12). "}
\end{quote}

This expert further explained that, in commercial art workflows, the primary difficulty in using AI is the imprecise control over the image features, such as lighting and texture.

\subsection{Scale to Spatial Control Tasks with GenAI}

The 3D interactive canvas allows users to directly manipulate the object in 3D to convey their spatial intentions to the generative model. 
In contrast to the current interactive system that requires an authoring process\cite{shi2025caring} or is constrained by predefined interactions \cite{he2023ubi}, our system can dynamically create interactive objects and environments based on the user prompt.
Additionally, using the spatial condition encoding function library, our system can translate the user's spatial intention to parametric spatial constraints compatible with the generative model. Therefore, it can scale to a variety of spatial manipulation tasks involving GenAI. For example, the user can use our system to create interactive geometry primitives to guide the generation of a 3D model \cite{dong2024coin3d}. 
The primitives can be placed as blocks to indicate the shape and affordance of the target object. 
Our system can also be used to deliver the human spatial instruction to the embedded AI, for example, indicating how the logistics items should be placed in a warehouse.







\section{Conclusion}

In this paper, we present \Oursystem{}, an image generation system that empowers users with precise spatial control. 
We achieve this by addressing two key considerations: (1) enabling user to precisely and intuitively express their spatial intents, and (2) accurately translating these intents into spatial constraints for the generative model.
The system provides users with an interactive 3D canvas, enabling them to express their intent by directly manipulating objects in a natural 3D manner. This interactive 3D canvas is dynamically generated by the system based on the user's prompt, thereby fulfilling the personalization requirements of the image generation task.
Once the user has set up the desired spatial composition for the image, the system uses a condition encoder to translate the user's spatial input into constraints for the generative model. 
Guided by these constraints, the generative model can accurately produce images that align with user expectations.
We conducted a two-session user study. In the comparative evaluation, all users preferred our system regarding interactivity, controllability, and ease of use. Quantitative result analysis also indicated that our system helps the generative model produce higher quality and more spatially accurate images. The system usability evaluation confirmed the performance of each module and resulted in a promising usability score for the overall system.
Therefore, we believe \Oursystem{} provides users with an easy and precise way to control the spatial factors to create images that closely match their spatial intents. 
We anticipate this work will inspire future research on controllable image generation and open new opportunities in other human-AI collaboration workflows, such as precise 3D generation and training or controlling embodied AI agents.

\bibliographystyle{ACM-Reference-Format}
\bibliography{ref}


\begin{thebibliography}{116}


\ifx \showCODEN    \undefined \def \showCODEN     #1{\unskip}     \fi
\ifx \showISBNx    \undefined \def \showISBNx     #1{\unskip}     \fi
\ifx \showISBNxiii \undefined \def \showISBNxiii  #1{\unskip}     \fi
\ifx \showISSN     \undefined \def \showISSN      #1{\unskip}     \fi
\ifx \showLCCN     \undefined \def \showLCCN      #1{\unskip}     \fi
\ifx \shownote     \undefined \def \shownote      #1{#1}          \fi
\ifx \showarticletitle \undefined \def \showarticletitle #1{#1}   \fi
\ifx \showURL      \undefined \def \showURL       {\relax}        \fi
\providecommand\bibfield[2]{#2}
\providecommand\bibinfo[2]{#2}
\providecommand\natexlab[1]{#1}
\providecommand\showeprint[2][]{arXiv:#2}

\bibitem[Alkemade et~al\mbox{.}(2017)]%
        {alkemade2017efficiency}
\bibfield{author}{\bibinfo{person}{Remi Alkemade}, \bibinfo{person}{Fons~J
  Verbeek}, {and} \bibinfo{person}{Stephan~G Lukosch}.}
  \bibinfo{year}{2017}\natexlab{}.
\newblock \showarticletitle{On the efficiency of a VR hand gesture-based
  interface for 3D object manipulations in conceptual design}.
\newblock \bibinfo{journal}{\emph{International Journal of Human--Computer
  Interaction}} \bibinfo{volume}{33}, \bibinfo{number}{11}
  (\bibinfo{year}{2017}), \bibinfo{pages}{882--901}.
\newblock


\bibitem[{Autodesk}(2025)]%
        {autodesk2025}
\bibfield{author}{\bibinfo{person}{{Autodesk}}.}
  \bibinfo{year}{2025}\natexlab{}.
\newblock \bibinfo{title}{{Autodesk | 3D Design, Engineering \& Construction
  Software}}.
\newblock
\urldef\tempurl%
\url{https://www.autodesk.com/}
\showURL{%
\tempurl}
\newblock
\shownote{Accessed: 2025-04-09}.


\bibitem[Avrahami et~al\mbox{.}(2024)]%
        {avrahami2024diffuhaul}
\bibfield{author}{\bibinfo{person}{Omri Avrahami}, \bibinfo{person}{Rinon Gal},
  \bibinfo{person}{Gal Chechik}, \bibinfo{person}{Ohad Fried},
  \bibinfo{person}{Dani Lischinski}, \bibinfo{person}{Arash Vahdat}, {and}
  \bibinfo{person}{Weili Nie}.} \bibinfo{year}{2024}\natexlab{}.
\newblock \showarticletitle{Diffuhaul: A training-free method for object
  dragging in images}. In \bibinfo{booktitle}{\emph{SIGGRAPH Asia 2024
  Conference Papers}}. \bibinfo{pages}{1--12}.
\newblock


\bibitem[Besan{\c{c}}on et~al\mbox{.}(2017)]%
        {besanccon2017mouse}
\bibfield{author}{\bibinfo{person}{Lonni Besan{\c{c}}on}, \bibinfo{person}{Paul
  Issartel}, \bibinfo{person}{Mehdi Ammi}, {and} \bibinfo{person}{Tobias
  Isenberg}.} \bibinfo{year}{2017}\natexlab{}.
\newblock \showarticletitle{Mouse, tactile, and tangible input for 3D
  manipulation}. In \bibinfo{booktitle}{\emph{Proceedings of the 2017 CHI
  conference on human factors in computing systems}}.
  \bibinfo{pages}{4727--4740}.
\newblock


\bibitem[Bhat et~al\mbox{.}(2024)]%
        {bhat2024loosecontrol}
\bibfield{author}{\bibinfo{person}{Shariq~Farooq Bhat}, \bibinfo{person}{Niloy
  Mitra}, {and} \bibinfo{person}{Peter Wonka}.}
  \bibinfo{year}{2024}\natexlab{}.
\newblock \showarticletitle{Loosecontrol: Lifting controlnet for generalized
  depth conditioning}. In \bibinfo{booktitle}{\emph{ACM SIGGRAPH 2024
  Conference Papers}}. \bibinfo{pages}{1--11}.
\newblock


\bibitem[Brade et~al\mbox{.}(2023)]%
        {brade2023promptify}
\bibfield{author}{\bibinfo{person}{Stephen Brade}, \bibinfo{person}{Bryan
  Wang}, \bibinfo{person}{Mauricio Sousa}, \bibinfo{person}{Sageev Oore}, {and}
  \bibinfo{person}{Tovi Grossman}.} \bibinfo{year}{2023}\natexlab{}.
\newblock \showarticletitle{Promptify: Text-to-image generation through
  interactive prompt exploration with large language models}. In
  \bibinfo{booktitle}{\emph{Proceedings of the 36th Annual ACM Symposium on
  User Interface Software and Technology}}. \bibinfo{pages}{1--14}.
\newblock


\bibitem[Cai et~al\mbox{.}(2024)]%
        {cai2024spatialbot}
\bibfield{author}{\bibinfo{person}{Wenxiao Cai}, \bibinfo{person}{Iaroslav
  Ponomarenko}, \bibinfo{person}{Jianhao Yuan}, \bibinfo{person}{Xiaoqi Li},
  \bibinfo{person}{Wankou Yang}, \bibinfo{person}{Hao Dong}, {and}
  \bibinfo{person}{Bo Zhao}.} \bibinfo{year}{2024}\natexlab{}.
\newblock \showarticletitle{Spatialbot: Precise spatial understanding with
  vision language models}.
\newblock \bibinfo{journal}{\emph{arXiv preprint arXiv:2406.13642}}
  (\bibinfo{year}{2024}).
\newblock


\bibitem[Cao et~al\mbox{.}(2019)]%
        {cao2019openpose}
\bibfield{author}{\bibinfo{person}{Zhe Cao}, \bibinfo{person}{Gines Hidalgo},
  \bibinfo{person}{Tomas Simon}, \bibinfo{person}{Shih-En Wei}, {and}
  \bibinfo{person}{Yaser Sheikh}.} \bibinfo{year}{2019}\natexlab{}.
\newblock \showarticletitle{Openpose: Realtime multi-person 2d pose estimation
  using part affinity fields}.
\newblock \bibinfo{journal}{\emph{IEEE transactions on pattern analysis and
  machine intelligence}} \bibinfo{volume}{43}, \bibinfo{number}{1}
  (\bibinfo{year}{2019}), \bibinfo{pages}{172--186}.
\newblock


\bibitem[Cao et~al\mbox{.}(2017)]%
        {cao2017realtime}
\bibfield{author}{\bibinfo{person}{Zhe Cao}, \bibinfo{person}{Tomas Simon},
  \bibinfo{person}{Shih-En Wei}, {and} \bibinfo{person}{Yaser Sheikh}.}
  \bibinfo{year}{2017}\natexlab{}.
\newblock \showarticletitle{Realtime multi-person 2d pose estimation using part
  affinity fields}. In \bibinfo{booktitle}{\emph{Proceedings of the IEEE
  conference on computer vision and pattern recognition}}.
  \bibinfo{pages}{7291--7299}.
\newblock


\bibitem[Chang et~al\mbox{.}(2015)]%
        {chang2015shapenet}
\bibfield{author}{\bibinfo{person}{Angel~X Chang}, \bibinfo{person}{Thomas
  Funkhouser}, \bibinfo{person}{Leonidas Guibas}, \bibinfo{person}{Pat
  Hanrahan}, \bibinfo{person}{Qixing Huang}, \bibinfo{person}{Zimo Li},
  \bibinfo{person}{Silvio Savarese}, \bibinfo{person}{Manolis Savva},
  \bibinfo{person}{Shuran Song}, \bibinfo{person}{Hao Su}, {et~al\mbox{.}}}
  \bibinfo{year}{2015}\natexlab{}.
\newblock \showarticletitle{Shapenet: An information-rich 3d model repository}.
\newblock \bibinfo{journal}{\emph{arXiv preprint arXiv:1512.03012}}
  (\bibinfo{year}{2015}).
\newblock


\bibitem[Chefer et~al\mbox{.}(2023)]%
        {chefer2023attend}
\bibfield{author}{\bibinfo{person}{Hila Chefer}, \bibinfo{person}{Yuval
  Alaluf}, \bibinfo{person}{Yael Vinker}, \bibinfo{person}{Lior Wolf}, {and}
  \bibinfo{person}{Daniel Cohen-Or}.} \bibinfo{year}{2023}\natexlab{}.
\newblock \showarticletitle{Attend-and-excite: Attention-based semantic
  guidance for text-to-image diffusion models}.
\newblock \bibinfo{journal}{\emph{ACM transactions on Graphics (TOG)}}
  \bibinfo{volume}{42}, \bibinfo{number}{4} (\bibinfo{year}{2023}),
  \bibinfo{pages}{1--10}.
\newblock


\bibitem[Choi et~al\mbox{.}(2024)]%
        {choi2024creativeconnect}
\bibfield{author}{\bibinfo{person}{DaEun Choi}, \bibinfo{person}{Sumin Hong},
  \bibinfo{person}{Jeongeon Park}, \bibinfo{person}{John Joon~Young Chung},
  {and} \bibinfo{person}{Juho Kim}.} \bibinfo{year}{2024}\natexlab{}.
\newblock \showarticletitle{CreativeConnect: Supporting Reference Recombination
  for Graphic Design Ideation with Generative AI}. In
  \bibinfo{booktitle}{\emph{Proceedings of the 2024 CHI Conference on Human
  Factors in Computing Systems}}. \bibinfo{pages}{1--25}.
\newblock


\bibitem[Chung and Adar(2023)]%
        {chung2023promptpaint}
\bibfield{author}{\bibinfo{person}{John Joon~Young Chung} {and}
  \bibinfo{person}{Eytan Adar}.} \bibinfo{year}{2023}\natexlab{}.
\newblock \showarticletitle{Promptpaint: Steering text-to-image generation
  through paint medium-like interactions}. In
  \bibinfo{booktitle}{\emph{Proceedings of the 36th Annual ACM Symposium on
  User Interface Software and Technology}}. \bibinfo{pages}{1--17}.
\newblock


\bibitem[Connor and Knierim(2017)]%
        {connor2017integration}
\bibfield{author}{\bibinfo{person}{Charles~E Connor} {and}
  \bibinfo{person}{James~J Knierim}.} \bibinfo{year}{2017}\natexlab{}.
\newblock \showarticletitle{Integration of objects and space in perception and
  memory}.
\newblock \bibinfo{journal}{\emph{Nature neuroscience}} \bibinfo{volume}{20},
  \bibinfo{number}{11} (\bibinfo{year}{2017}), \bibinfo{pages}{1493--1503}.
\newblock


\bibitem[Dang et~al\mbox{.}(2022)]%
        {dang2022ganslider}
\bibfield{author}{\bibinfo{person}{Hai Dang}, \bibinfo{person}{Lukas Mecke},
  {and} \bibinfo{person}{Daniel Buschek}.} \bibinfo{year}{2022}\natexlab{}.
\newblock \showarticletitle{Ganslider: How users control generative models for
  images using multiple sliders with and without feedforward information}. In
  \bibinfo{booktitle}{\emph{Proceedings of the 2022 CHI Conference on Human
  Factors in Computing Systems}}. \bibinfo{pages}{1--15}.
\newblock


\bibitem[Degrave et~al\mbox{.}(2019)]%
        {degrave2019differentiable}
\bibfield{author}{\bibinfo{person}{Jonas Degrave}, \bibinfo{person}{Michiel
  Hermans}, \bibinfo{person}{Joni Dambre}, {and} \bibinfo{person}{Francis
  Wyffels}.} \bibinfo{year}{2019}\natexlab{}.
\newblock \showarticletitle{A differentiable physics engine for deep learning
  in robotics}.
\newblock \bibinfo{journal}{\emph{Frontiers in neurorobotics}}
  \bibinfo{volume}{13} (\bibinfo{year}{2019}), \bibinfo{pages}{6}.
\newblock


\bibitem[Deitke et~al\mbox{.}(2023)]%
        {deitke2023objaverse}
\bibfield{author}{\bibinfo{person}{Matt Deitke}, \bibinfo{person}{Dustin
  Schwenk}, \bibinfo{person}{Jordi Salvador}, \bibinfo{person}{Luca Weihs},
  \bibinfo{person}{Oscar Michel}, \bibinfo{person}{Eli VanderBilt},
  \bibinfo{person}{Ludwig Schmidt}, \bibinfo{person}{Kiana Ehsani},
  \bibinfo{person}{Aniruddha Kembhavi}, {and} \bibinfo{person}{Ali Farhadi}.}
  \bibinfo{year}{2023}\natexlab{}.
\newblock \showarticletitle{Objaverse: A universe of annotated 3d objects}. In
  \bibinfo{booktitle}{\emph{Proceedings of the IEEE/CVF conference on computer
  vision and pattern recognition}}. \bibinfo{pages}{13142--13153}.
\newblock


\bibitem[Dolins and Mitchell(2010)]%
        {dolins2010spatial}
\bibfield{author}{\bibinfo{person}{Francine~L Dolins} {and}
  \bibinfo{person}{Robert~W Mitchell}.} \bibinfo{year}{2010}\natexlab{}.
\newblock \bibinfo{booktitle}{\emph{Spatial cognition, spatial perception:
  mapping the self and space}}.
\newblock \bibinfo{publisher}{Cambridge University Press}.
\newblock


\bibitem[Dong et~al\mbox{.}(2024)]%
        {dong2024coin3d}
\bibfield{author}{\bibinfo{person}{Wenqi Dong}, \bibinfo{person}{Bangbang
  Yang}, \bibinfo{person}{Lin Ma}, \bibinfo{person}{Xiao Liu},
  \bibinfo{person}{Liyuan Cui}, \bibinfo{person}{Hujun Bao},
  \bibinfo{person}{Yuewen Ma}, {and} \bibinfo{person}{Zhaopeng Cui}.}
  \bibinfo{year}{2024}\natexlab{}.
\newblock \showarticletitle{Coin3d: Controllable and interactive 3d assets
  generation with proxy-guided conditioning}. In \bibinfo{booktitle}{\emph{ACM
  SIGGRAPH 2024 Conference Papers}}. \bibinfo{pages}{1--10}.
\newblock


\bibitem[Duan et~al\mbox{.}(2025)]%
        {duan2025parametric}
\bibfield{author}{\bibinfo{person}{Runlin Duan}, \bibinfo{person}{Xiyun Hu},
  \bibinfo{person}{Min Liu}, \bibinfo{person}{Jingyu Shi}, {and}
  \bibinfo{person}{Karthik Ramani}.} \bibinfo{year}{2025}\natexlab{}.
\newblock \showarticletitle{pARametric: Empowering In Situ Parametric Modeling
  in Augment Reality for Personal Fabrication}.
\newblock \bibinfo{journal}{\emph{Journal of Computing and Information Science
  in Engineering}} \bibinfo{volume}{25}, \bibinfo{number}{4}
  (\bibinfo{year}{2025}), \bibinfo{pages}{041001}.
\newblock


\bibitem[Duan et~al\mbox{.}(2024)]%
        {duan2024conceptvis}
\bibfield{author}{\bibinfo{person}{Runlin Duan}, \bibinfo{person}{Nachiketh
  Karthik}, \bibinfo{person}{Jingyu Shi}, \bibinfo{person}{Rahul Jain},
  \bibinfo{person}{Maria~C Yang}, {and} \bibinfo{person}{Karthik Ramani}.}
  \bibinfo{year}{2024}\natexlab{}.
\newblock \showarticletitle{ConceptVis: Generating and Exploring Design
  Concepts for Early-Stage Ideation Using Large Language Model}. In
  \bibinfo{booktitle}{\emph{International Design Engineering Technical
  Conferences and Computers and Information in Engineering Conference}},
  Vol.~\bibinfo{volume}{88377}. American Society of Mechanical Engineers,
  \bibinfo{pages}{V03BT03A042}.
\newblock


\bibitem[Eberly(2006)]%
        {eberly20063d}
\bibfield{author}{\bibinfo{person}{David Eberly}.}
  \bibinfo{year}{2006}\natexlab{}.
\newblock \bibinfo{booktitle}{\emph{3D game engine design: a practical approach
  to real-time computer graphics}}.
\newblock \bibinfo{publisher}{CRC Press}.
\newblock


\bibitem[Eldesokey and Wonka(2024)]%
        {eldesokey2024build}
\bibfield{author}{\bibinfo{person}{Abdelrahman Eldesokey} {and}
  \bibinfo{person}{Peter Wonka}.} \bibinfo{year}{2024}\natexlab{}.
\newblock \showarticletitle{Build-a-scene: Interactive 3d layout control for
  diffusion-based image generation}.
\newblock \bibinfo{journal}{\emph{arXiv preprint arXiv:2408.14819}}
  (\bibinfo{year}{2024}).
\newblock


\bibitem[Exocortex~Technologies(2013)]%
        {claraio}
\bibfield{author}{\bibinfo{person}{Inc. Exocortex~Technologies}.}
  \bibinfo{year}{2013}\natexlab{}.
\newblock \bibinfo{title}{Clara.io: Online 3D Modeling, 3D Rendering, Free 3D
  Models}.
\newblock
\urldef\tempurl%
\url{https://clara.io/}
\showURL{%
\tempurl}
\newblock
\shownote{Accessed: 2025-04-07}.


\bibitem[Fairbrother(1974)]%
        {fairbrother1974nature}
\bibfield{author}{\bibinfo{person}{Nan Fairbrother}.}
  \bibinfo{year}{1974}\natexlab{}.
\newblock \bibinfo{booktitle}{\emph{The nature of landscape design}}.
\newblock \bibinfo{publisher}{Architectural Press London}.
\newblock


\bibitem[Feng et~al\mbox{.}(2023)]%
        {feng2023layoutgpt}
\bibfield{author}{\bibinfo{person}{Weixi Feng}, \bibinfo{person}{Wanrong Zhu},
  \bibinfo{person}{Tsu-jui Fu}, \bibinfo{person}{Varun Jampani},
  \bibinfo{person}{Arjun Akula}, \bibinfo{person}{Xuehai He},
  \bibinfo{person}{Sugato Basu}, \bibinfo{person}{Xin~Eric Wang}, {and}
  \bibinfo{person}{William~Yang Wang}.} \bibinfo{year}{2023}\natexlab{}.
\newblock \showarticletitle{Layoutgpt: Compositional visual planning and
  generation with large language models}.
\newblock \bibinfo{journal}{\emph{Advances in Neural Information Processing
  Systems}}  \bibinfo{volume}{36} (\bibinfo{year}{2023}),
  \bibinfo{pages}{18225--18250}.
\newblock


\bibitem[Fiorella et~al\mbox{.}(2010)]%
        {fiorella2010multi}
\bibfield{author}{\bibinfo{person}{Donato Fiorella}, \bibinfo{person}{Andrea
  Sanna}, {and} \bibinfo{person}{Fabrizio Lamberti}.}
  \bibinfo{year}{2010}\natexlab{}.
\newblock \showarticletitle{Multi-touch user interface evaluation for 3D object
  manipulation on mobile devices}.
\newblock \bibinfo{journal}{\emph{Journal on Multimodal User Interfaces}}
  \bibinfo{volume}{4} (\bibinfo{year}{2010}), \bibinfo{pages}{3--10}.
\newblock


\bibitem[Galati et~al\mbox{.}(2010)]%
        {galati2010multiple}
\bibfield{author}{\bibinfo{person}{Gaspare Galati}, \bibinfo{person}{Gina
  Pelle}, \bibinfo{person}{Alain Berthoz}, {and} \bibinfo{person}{Giorgia
  Committeri}.} \bibinfo{year}{2010}\natexlab{}.
\newblock \showarticletitle{Multiple reference frames used by the human brain
  for spatial perception and memory}.
\newblock \bibinfo{journal}{\emph{Experimental brain research}}
  \bibinfo{volume}{206} (\bibinfo{year}{2010}), \bibinfo{pages}{109--120}.
\newblock


\bibitem[Gao et~al\mbox{.}(2024)]%
        {gao2024cat3d}
\bibfield{author}{\bibinfo{person}{Ruiqi Gao}, \bibinfo{person}{Aleksander
  Holynski}, \bibinfo{person}{Philipp Henzler}, \bibinfo{person}{Arthur
  Brussee}, \bibinfo{person}{Ricardo Martin-Brualla}, \bibinfo{person}{Pratul
  Srinivasan}, \bibinfo{person}{Jonathan~T Barron}, {and} \bibinfo{person}{Ben
  Poole}.} \bibinfo{year}{2024}\natexlab{}.
\newblock \showarticletitle{Cat3d: Create anything in 3d with multi-view
  diffusion models}.
\newblock \bibinfo{journal}{\emph{arXiv preprint arXiv:2405.10314}}
  (\bibinfo{year}{2024}).
\newblock


\bibitem[Gibson(2014)]%
        {gibson2014ecological}
\bibfield{author}{\bibinfo{person}{James~J Gibson}.}
  \bibinfo{year}{2014}\natexlab{}.
\newblock \bibinfo{booktitle}{\emph{The ecological approach to visual
  perception: classic edition}}.
\newblock \bibinfo{publisher}{Psychology press}.
\newblock


\bibitem[Goh et~al\mbox{.}(2019)]%
        {goh20193d}
\bibfield{author}{\bibinfo{person}{Eg~Su Goh}, \bibinfo{person}{Mohd~Shahrizal
  Sunar}, {and} \bibinfo{person}{Ajune~Wanis Ismail}.}
  \bibinfo{year}{2019}\natexlab{}.
\newblock \showarticletitle{3D object manipulation techniques in handheld
  mobile augmented reality interface: A review}.
\newblock \bibinfo{journal}{\emph{IEEE Access}}  \bibinfo{volume}{7}
  (\bibinfo{year}{2019}), \bibinfo{pages}{40581--40601}.
\newblock


\bibitem[Gokhale et~al\mbox{.}(2022)]%
        {gokhale2022benchmarking}
\bibfield{author}{\bibinfo{person}{Tejas Gokhale}, \bibinfo{person}{Hamid
  Palangi}, \bibinfo{person}{Besmira Nushi}, \bibinfo{person}{Vibhav Vineet},
  \bibinfo{person}{Eric Horvitz}, \bibinfo{person}{Ece Kamar},
  \bibinfo{person}{Chitta Baral}, {and} \bibinfo{person}{Yezhou Yang}.}
  \bibinfo{year}{2022}\natexlab{}.
\newblock \showarticletitle{Benchmarking spatial relationships in text-to-image
  generation}.
\newblock \bibinfo{journal}{\emph{arXiv preprint arXiv:2212.10015}}
  (\bibinfo{year}{2022}).
\newblock


\bibitem[Gooch et~al\mbox{.}(2001)]%
        {gooch2001artistic}
\bibfield{author}{\bibinfo{person}{Bruce Gooch}, \bibinfo{person}{Erik
  Reinhard}, \bibinfo{person}{Chris Moulding}, {and} \bibinfo{person}{Peter
  Shirley}.} \bibinfo{year}{2001}\natexlab{}.
\newblock \showarticletitle{Artistic composition for image creation}. In
  \bibinfo{booktitle}{\emph{Rendering Techniques 2001: Proceedings of the
  Eurographics Workshop in London, United Kingdom, June 25--27, 2001 12}}.
  Springer, \bibinfo{pages}{83--88}.
\newblock


\bibitem[Gopher et~al\mbox{.}(2000)]%
        {gopher2000switching}
\bibfield{author}{\bibinfo{person}{Daniel Gopher}, \bibinfo{person}{Lilach
  Armony}, {and} \bibinfo{person}{Yaakov Greenshpan}.}
  \bibinfo{year}{2000}\natexlab{}.
\newblock \showarticletitle{Switching tasks and attention policies.}
\newblock \bibinfo{journal}{\emph{Journal of Experimental Psychology: General}}
  \bibinfo{volume}{129}, \bibinfo{number}{3} (\bibinfo{year}{2000}),
  \bibinfo{pages}{308}.
\newblock


\bibitem[Graham and Redies(2010)]%
        {graham2010statistical}
\bibfield{author}{\bibinfo{person}{Daniel~J Graham} {and}
  \bibinfo{person}{Christoph Redies}.} \bibinfo{year}{2010}\natexlab{}.
\newblock \showarticletitle{Statistical regularities in art: Relations with
  visual coding and perception}.
\newblock \bibinfo{journal}{\emph{Vision research}} \bibinfo{volume}{50},
  \bibinfo{number}{16} (\bibinfo{year}{2010}), \bibinfo{pages}{1503--1509}.
\newblock


\bibitem[Hart(1988)]%
        {hart1988development}
\bibfield{author}{\bibinfo{person}{SG Hart}.} \bibinfo{year}{1988}\natexlab{}.
\newblock \showarticletitle{Development of NASA-TLX (Task Load Index): Results
  of empirical and theoretical research}.
\newblock \bibinfo{journal}{\emph{Human mental workload/Elsevier}}
  (\bibinfo{year}{1988}).
\newblock


\bibitem[He et~al\mbox{.}(2023)]%
        {he2023ubi}
\bibfield{author}{\bibinfo{person}{Fengming He}, \bibinfo{person}{Xiyun Hu},
  \bibinfo{person}{Jingyu Shi}, \bibinfo{person}{Xun Qian},
  \bibinfo{person}{Tianyi Wang}, {and} \bibinfo{person}{Karthik Ramani}.}
  \bibinfo{year}{2023}\natexlab{}.
\newblock \showarticletitle{UBI edge: authoring edge-based opportunistic
  tangible user interfaces in augmented reality}. In
  \bibinfo{booktitle}{\emph{Proceedings of the 2023 CHI Conference on Human
  Factors in Computing Systems}}. \bibinfo{pages}{1--14}.
\newblock


\bibitem[Hou et~al\mbox{.}(2024)]%
        {hou2024c2ideas}
\bibfield{author}{\bibinfo{person}{Yihan Hou}, \bibinfo{person}{Manling Yang},
  \bibinfo{person}{Hao Cui}, \bibinfo{person}{Lei Wang}, \bibinfo{person}{Jie
  Xu}, {and} \bibinfo{person}{Wei Zeng}.} \bibinfo{year}{2024}\natexlab{}.
\newblock \showarticletitle{C2Ideas: Supporting Creative Interior Color Design
  Ideation with a Large Language Model}. In
  \bibinfo{booktitle}{\emph{Proceedings of the CHI Conference on Human Factors
  in Computing Systems}}. \bibinfo{pages}{1--18}.
\newblock


\bibitem[Huang et~al\mbox{.}(2025)]%
        {huang2025t2i}
\bibfield{author}{\bibinfo{person}{Kaiyi Huang}, \bibinfo{person}{Chengqi
  Duan}, \bibinfo{person}{Kaiyue Sun}, \bibinfo{person}{Enze Xie},
  \bibinfo{person}{Zhenguo Li}, {and} \bibinfo{person}{Xihui Liu}.}
  \bibinfo{year}{2025}\natexlab{}.
\newblock \showarticletitle{T2I-CompBench++: An Enhanced and Comprehensive
  Benchmark for Compositional Text-to-Image Generation}.
\newblock \bibinfo{journal}{\emph{IEEE Transactions on Pattern Analysis and
  Machine Intelligence}} (\bibinfo{year}{2025}).
\newblock


\bibitem[Huang et~al\mbox{.}(2018)]%
        {huang2018light}
\bibfield{author}{\bibinfo{person}{Zheng Huang}, \bibinfo{person}{Qiang Liu},
  \bibinfo{person}{Stephen Westland}, \bibinfo{person}{Michael~R Pointer},
  \bibinfo{person}{M~Ronnier Luo}, {and} \bibinfo{person}{Kaida Xiao}.}
  \bibinfo{year}{2018}\natexlab{}.
\newblock \showarticletitle{Light dominates colour preference when correlated
  colour temperature differs}.
\newblock \bibinfo{journal}{\emph{Lighting Research \& Technology}}
  \bibinfo{volume}{50}, \bibinfo{number}{7} (\bibinfo{year}{2018}),
  \bibinfo{pages}{995--1012}.
\newblock


\bibitem[Inc.(2025)]%
        {substance3dstager}
\bibfield{author}{\bibinfo{person}{Adobe Inc.}}
  \bibinfo{year}{2025}\natexlab{}.
\newblock \bibinfo{title}{Adobe Substance 3D Stager}.
\newblock
\urldef\tempurl%
\url{https://www.adobe.com/products/substance3d/apps/stager.html}
\showURL{%
\tempurl}
\newblock
\shownote{Accessed: 2025-04-07}.


\bibitem[Jankowski and Hachet(2013)]%
        {jankowski2013survey}
\bibfield{author}{\bibinfo{person}{Jacek Jankowski} {and}
  \bibinfo{person}{Martin Hachet}.} \bibinfo{year}{2013}\natexlab{}.
\newblock \showarticletitle{A survey of interaction techniques for interactive
  3D environments}. In \bibinfo{booktitle}{\emph{Eurographics 2013-STAR}}.
\newblock


\bibitem[Jeon et~al\mbox{.}(2021)]%
        {jeon2021fashionq}
\bibfield{author}{\bibinfo{person}{Youngseung Jeon}, \bibinfo{person}{Seungwan
  Jin}, \bibinfo{person}{Patrick~C Shih}, {and} \bibinfo{person}{Kyungsik
  Han}.} \bibinfo{year}{2021}\natexlab{}.
\newblock \showarticletitle{FashionQ: an ai-driven creativity support tool for
  facilitating ideation in fashion design}. In
  \bibinfo{booktitle}{\emph{Proceedings of the 2021 CHI Conference on Human
  Factors in Computing Systems}}. \bibinfo{pages}{1--18}.
\newblock


\bibitem[Ju et~al\mbox{.}(2023)]%
        {ju2023humansd}
\bibfield{author}{\bibinfo{person}{Xuan Ju}, \bibinfo{person}{Ailing Zeng},
  \bibinfo{person}{Chenchen Zhao}, \bibinfo{person}{Jianan Wang},
  \bibinfo{person}{Lei Zhang}, {and} \bibinfo{person}{Qiang Xu}.}
  \bibinfo{year}{2023}\natexlab{}.
\newblock \showarticletitle{Humansd: A native skeleton-guided diffusion model
  for human image generation}. In \bibinfo{booktitle}{\emph{Proceedings of the
  IEEE/CVF International Conference on Computer Vision}}.
  \bibinfo{pages}{15988--15998}.
\newblock


\bibitem[Khan and Vogel(2012)]%
        {khan2012evaluating}
\bibfield{author}{\bibinfo{person}{Shehroz~S Khan} {and}
  \bibinfo{person}{Daniel Vogel}.} \bibinfo{year}{2012}\natexlab{}.
\newblock \showarticletitle{Evaluating visual aesthetics in photographic
  portraiture}. In \bibinfo{booktitle}{\emph{Proceedings of the Eighth Annual
  Symposium on Computational Aesthetics in Graphics, visualization, and
  imaging}}. \bibinfo{pages}{55--62}.
\newblock


\bibitem[Kim and Lee(2016)]%
        {kim2016touch}
\bibfield{author}{\bibinfo{person}{Minseok Kim} {and} \bibinfo{person}{Jae~Yeol
  Lee}.} \bibinfo{year}{2016}\natexlab{}.
\newblock \showarticletitle{Touch and hand gesture-based interactions for
  directly manipulating 3D virtual objects in mobile augmented reality}.
\newblock \bibinfo{journal}{\emph{Multimedia Tools and Applications}}
  \bibinfo{volume}{75} (\bibinfo{year}{2016}), \bibinfo{pages}{16529--16550}.
\newblock


\bibitem[Koch et~al\mbox{.}(2019)]%
        {koch2019may}
\bibfield{author}{\bibinfo{person}{Janin Koch}, \bibinfo{person}{Andr{\'e}s
  Lucero}, \bibinfo{person}{Lena Hegemann}, {and} \bibinfo{person}{Antti
  Oulasvirta}.} \bibinfo{year}{2019}\natexlab{}.
\newblock \showarticletitle{May AI? Design ideation with cooperative contextual
  bandits}. In \bibinfo{booktitle}{\emph{Proceedings of the 2019 CHI Conference
  on Human Factors in Computing Systems}}. \bibinfo{pages}{1--12}.
\newblock


\bibitem[Leonhardt et~al\mbox{.}(2015)]%
        {leonhardt2015your}
\bibfield{author}{\bibinfo{person}{James~M Leonhardt}, \bibinfo{person}{Jesse~R
  Catlin}, {and} \bibinfo{person}{Dante~M Pirouz}.}
  \bibinfo{year}{2015}\natexlab{}.
\newblock \showarticletitle{Is your product facing the ad's center? Facing
  direction affects processing fluency and ad evaluation}.
\newblock \bibinfo{journal}{\emph{Journal of Advertising}}
  \bibinfo{volume}{44}, \bibinfo{number}{4} (\bibinfo{year}{2015}),
  \bibinfo{pages}{315--325}.
\newblock


\bibitem[Leyssen et~al\mbox{.}(2012)]%
        {leyssen2012aesthetic}
\bibfield{author}{\bibinfo{person}{Mieke~HR Leyssen}, \bibinfo{person}{Sarah
  Linsen}, \bibinfo{person}{Jonathan Sammartino}, {and}
  \bibinfo{person}{Stephen~E Palmer}.} \bibinfo{year}{2012}\natexlab{}.
\newblock \showarticletitle{Aesthetic preference for spatial composition in
  multiobject pictures}.
\newblock \bibinfo{journal}{\emph{i-Perception}} \bibinfo{volume}{3},
  \bibinfo{number}{1} (\bibinfo{year}{2012}), \bibinfo{pages}{25--49}.
\newblock


\bibitem[Li et~al\mbox{.}(2024b)]%
        {li2024anicraft}
\bibfield{author}{\bibinfo{person}{Boyu Li}, \bibinfo{person}{Linping Yuan},
  \bibinfo{person}{Zhe Yan}, \bibinfo{person}{Qianxi Liu},
  \bibinfo{person}{Yulin Shen}, {and} \bibinfo{person}{Zeyu Wang}.}
  \bibinfo{year}{2024}\natexlab{b}.
\newblock \showarticletitle{AniCraft: Crafting Everyday Objects as Physical
  Proxies for Prototyping 3D Character Animation in Mixed Reality}. In
  \bibinfo{booktitle}{\emph{Proceedings of the 37th Annual ACM Symposium on
  User Interface Software and Technology}}. \bibinfo{pages}{1--14}.
\newblock


\bibitem[Li et~al\mbox{.}(2024a)]%
        {li2024simple}
\bibfield{author}{\bibinfo{person}{Xirui Li}, \bibinfo{person}{Charles
  Herrmann}, \bibinfo{person}{Kelvin~CK Chan}, \bibinfo{person}{Yinxiao Li},
  \bibinfo{person}{Deqing Sun}, \bibinfo{person}{Chao Ma}, {and}
  \bibinfo{person}{Ming-Hsuan Yang}.} \bibinfo{year}{2024}\natexlab{a}.
\newblock \showarticletitle{A simple approach to unifying diffusion-based
  conditional generation}.
\newblock \bibinfo{journal}{\emph{arXiv preprint arXiv:2410.11439}}
  (\bibinfo{year}{2024}).
\newblock


\bibitem[Liao et~al\mbox{.}(2022)]%
        {liao2022text}
\bibfield{author}{\bibinfo{person}{Wentong Liao}, \bibinfo{person}{Kai Hu},
  \bibinfo{person}{Michael~Ying Yang}, {and} \bibinfo{person}{Bodo Rosenhahn}.}
  \bibinfo{year}{2022}\natexlab{}.
\newblock \showarticletitle{Text to image generation with semantic-spatial
  aware gan}. In \bibinfo{booktitle}{\emph{Proceedings of the IEEE/CVF
  conference on computer vision and pattern recognition}}.
  \bibinfo{pages}{18187--18196}.
\newblock


\bibitem[Lin et~al\mbox{.}(2024)]%
        {lin2024inkspire}
\bibfield{author}{\bibinfo{person}{David Chuan-En Lin},
  \bibinfo{person}{Hyeonsu~B Kang}, \bibinfo{person}{Nikolas Martelaro},
  \bibinfo{person}{Aniket Kittur}, \bibinfo{person}{Yan-Ying Chen}, {and}
  \bibinfo{person}{Matthew~K Hong}.} \bibinfo{year}{2024}\natexlab{}.
\newblock \showarticletitle{Inkspire: Sketching Product Designs with AI}. In
  \bibinfo{booktitle}{\emph{Adjunct Proceedings of the 37th Annual ACM
  Symposium on User Interface Software and Technology}}. \bibinfo{pages}{1--6}.
\newblock


\bibitem[Lin et~al\mbox{.}(2025)]%
        {lin2025inkspire}
\bibfield{author}{\bibinfo{person}{David Chuan-En Lin},
  \bibinfo{person}{Hyeonsu~B Kang}, \bibinfo{person}{Nikolas Martelaro},
  \bibinfo{person}{Aniket Kittur}, \bibinfo{person}{Yan-Ying Chen}, {and}
  \bibinfo{person}{Matthew~K Hong}.} \bibinfo{year}{2025}\natexlab{}.
\newblock \showarticletitle{Inkspire: Supporting Design Exploration with
  Generative AI through Analogical Sketching}.
\newblock \bibinfo{journal}{\emph{arXiv preprint arXiv:2501.18588}}
  (\bibinfo{year}{2025}).
\newblock


\bibitem[Liu et~al\mbox{.}(2025)]%
        {liu2025generative}
\bibfield{author}{\bibinfo{person}{Daochang Liu}, \bibinfo{person}{Junyu
  Zhang}, \bibinfo{person}{Anh-Dung Dinh}, \bibinfo{person}{Eunbyung Park},
  \bibinfo{person}{Shichao Zhang}, {and} \bibinfo{person}{Chang Xu}.}
  \bibinfo{year}{2025}\natexlab{}.
\newblock \showarticletitle{Generative Physical AI in Vision: A Survey}.
\newblock \bibinfo{journal}{\emph{arXiv preprint arXiv:2501.10928}}
  (\bibinfo{year}{2025}).
\newblock


\bibitem[Liu et~al\mbox{.}(2023a)]%
        {liu20233dall}
\bibfield{author}{\bibinfo{person}{Vivian Liu}, \bibinfo{person}{Jo Vermeulen},
  \bibinfo{person}{George Fitzmaurice}, {and} \bibinfo{person}{Justin
  Matejka}.} \bibinfo{year}{2023}\natexlab{a}.
\newblock \showarticletitle{3DALL-E: Integrating text-to-image AI in 3D design
  workflows}. In \bibinfo{booktitle}{\emph{Proceedings of the 2023 ACM
  designing interactive systems conference}}. \bibinfo{pages}{1955--1977}.
\newblock


\bibitem[Liu et~al\mbox{.}(2023b)]%
        {liu2023instrumentar}
\bibfield{author}{\bibinfo{person}{Ziyi Liu}, \bibinfo{person}{Zhengzhe Zhu},
  \bibinfo{person}{Enze Jiang}, \bibinfo{person}{Feichi Huang},
  \bibinfo{person}{Ana~M Villanueva}, \bibinfo{person}{Xun Qian},
  \bibinfo{person}{Tianyi Wang}, {and} \bibinfo{person}{Karthik Ramani}.}
  \bibinfo{year}{2023}\natexlab{b}.
\newblock \showarticletitle{Instrumentar: Auto-generation of augmented reality
  tutorials for operating digital instruments through recording embodied
  demonstration}. In \bibinfo{booktitle}{\emph{Proceedings of the 2023 CHI
  Conference on Human Factors in Computing Systems}}. \bibinfo{pages}{1--17}.
\newblock


\bibitem[Luo et~al\mbox{.}(2023)]%
        {luo2023scalable}
\bibfield{author}{\bibinfo{person}{Tiange Luo}, \bibinfo{person}{Chris
  Rockwell}, \bibinfo{person}{Honglak Lee}, {and} \bibinfo{person}{Justin
  Johnson}.} \bibinfo{year}{2023}\natexlab{}.
\newblock \showarticletitle{Scalable 3d captioning with pretrained models}.
\newblock \bibinfo{journal}{\emph{Advances in Neural Information Processing
  Systems}}  \bibinfo{volume}{36} (\bibinfo{year}{2023}),
  \bibinfo{pages}{75307--75337}.
\newblock


\bibitem[Maqbool(2023)]%
        {maqbool2023aesthetic}
\bibfield{author}{\bibinfo{person}{Hira Maqbool}.}
  \bibinfo{year}{2023}\natexlab{}.
\newblock \showarticletitle{Aesthetic choices: Defining the range of aesthetic
  views in interactive digital media including games and 3D virtual
  environments (3D VEs)}.
\newblock  (\bibinfo{year}{2023}).
\newblock


\bibitem[Mendes et~al\mbox{.}(2019)]%
        {mendes2019survey}
\bibfield{author}{\bibinfo{person}{Daniel Mendes}, \bibinfo{person}{Fabio~Marco
  Caputo}, \bibinfo{person}{Andrea Giachetti}, \bibinfo{person}{Alfredo
  Ferreira}, {and} \bibinfo{person}{Joaquim Jorge}.}
  \bibinfo{year}{2019}\natexlab{}.
\newblock \showarticletitle{A survey on 3d virtual object manipulation: From
  the desktop to immersive virtual environments}. In
  \bibinfo{booktitle}{\emph{Computer graphics forum}},
  Vol.~\bibinfo{volume}{38}. Wiley Online Library, \bibinfo{pages}{21--45}.
\newblock


\bibitem[Meng et~al\mbox{.}(2025)]%
        {meng2025grounding}
\bibfield{author}{\bibinfo{person}{Siwei Meng}, \bibinfo{person}{Yawei Luo},
  {and} \bibinfo{person}{Ping Liu}.} \bibinfo{year}{2025}\natexlab{}.
\newblock \showarticletitle{Grounding Creativity in Physics: A Brief Survey of
  Physical Priors in AIGC}.
\newblock \bibinfo{journal}{\emph{arXiv preprint arXiv:2502.07007}}
  (\bibinfo{year}{2025}).
\newblock


\bibitem[Mi et~al\mbox{.}(2025)]%
        {mi2025think}
\bibfield{author}{\bibinfo{person}{Zhenxing Mi}, \bibinfo{person}{Kuan-Chieh
  Wang}, \bibinfo{person}{Guocheng Qian}, \bibinfo{person}{Hanrong Ye},
  \bibinfo{person}{Runtao Liu}, \bibinfo{person}{Sergey Tulyakov},
  \bibinfo{person}{Kfir Aberman}, {and} \bibinfo{person}{Dan Xu}.}
  \bibinfo{year}{2025}\natexlab{}.
\newblock \showarticletitle{I Think, Therefore I Diffuse: Enabling Multimodal
  In-Context Reasoning in Diffusion Models}.
\newblock \bibinfo{journal}{\emph{arXiv preprint arXiv:2502.10458}}
  (\bibinfo{year}{2025}).
\newblock


\bibitem[Mo et~al\mbox{.}(2024)]%
        {mo2024freecontrol}
\bibfield{author}{\bibinfo{person}{Sicheng Mo}, \bibinfo{person}{Fangzhou Mu},
  \bibinfo{person}{Kuan~Heng Lin}, \bibinfo{person}{Yanli Liu},
  \bibinfo{person}{Bochen Guan}, \bibinfo{person}{Yin Li}, {and}
  \bibinfo{person}{Bolei Zhou}.} \bibinfo{year}{2024}\natexlab{}.
\newblock \showarticletitle{Freecontrol: Training-free spatial control of any
  text-to-image diffusion model with any condition}. In
  \bibinfo{booktitle}{\emph{Proceedings of the IEEE/CVF Conference on Computer
  Vision and Pattern Recognition}}. \bibinfo{pages}{7465--7475}.
\newblock


\bibitem[Mou et~al\mbox{.}(2024)]%
        {mou2024t2i}
\bibfield{author}{\bibinfo{person}{Chong Mou}, \bibinfo{person}{Xintao Wang},
  \bibinfo{person}{Liangbin Xie}, \bibinfo{person}{Yanze Wu},
  \bibinfo{person}{Jian Zhang}, \bibinfo{person}{Zhongang Qi}, {and}
  \bibinfo{person}{Ying Shan}.} \bibinfo{year}{2024}\natexlab{}.
\newblock \showarticletitle{T2i-adapter: Learning adapters to dig out more
  controllable ability for text-to-image diffusion models}. In
  \bibinfo{booktitle}{\emph{Proceedings of the AAAI conference on artificial
  intelligence}}, Vol.~\bibinfo{volume}{38}. \bibinfo{pages}{4296--4304}.
\newblock


\bibitem[Neroni et~al\mbox{.}(2021)]%
        {neroni2021virtual}
\bibfield{author}{\bibinfo{person}{Maria~Adriana Neroni},
  \bibinfo{person}{Alfred Oti}, {and} \bibinfo{person}{Nathan Crilly}.}
  \bibinfo{year}{2021}\natexlab{}.
\newblock \showarticletitle{Virtual Reality design-build-test games with
  physics simulation: opportunities for researching design cognition}.
\newblock \bibinfo{journal}{\emph{International Journal of Design Creativity
  and Innovation}} \bibinfo{volume}{9}, \bibinfo{number}{3}
  (\bibinfo{year}{2021}), \bibinfo{pages}{139--173}.
\newblock


\bibitem[Oh et~al\mbox{.}(2024)]%
        {oh2024lumimood}
\bibfield{author}{\bibinfo{person}{Jeongseok Oh}, \bibinfo{person}{Seungju
  Kim}, {and} \bibinfo{person}{Seungjun Kim}.} \bibinfo{year}{2024}\natexlab{}.
\newblock \showarticletitle{LumiMood: A Creativity Support Tool for Designing
  the Mood of a 3D Scene}. In \bibinfo{booktitle}{\emph{Proceedings of the 2024
  CHI Conference on Human Factors in Computing Systems}}.
  \bibinfo{pages}{1--21}.
\newblock


\bibitem[{Open Robotics}(2025)]%
        {ros2025}
\bibfield{author}{\bibinfo{person}{{Open Robotics}}.}
  \bibinfo{year}{2025}\natexlab{}.
\newblock \bibinfo{title}{{Robot Operating System (ROS)}}.
\newblock
\urldef\tempurl%
\url{https://www.ros.org/}
\showURL{%
\tempurl}
\newblock
\shownote{Accessed: 2025-04-09}.


\bibitem[OpenAI(2025)]%
        {openai}
\bibfield{author}{\bibinfo{person}{OpenAI}.} \bibinfo{year}{2025}\natexlab{}.
\newblock \bibinfo{title}{OpenAI}.
\newblock
\urldef\tempurl%
\url{https://openai.com/}
\showURL{%
\tempurl}
\newblock
\shownote{Accessed: 2025-04-09}.


\bibitem[Oppenlaender et~al\mbox{.}(2024)]%
        {oppenlaender2024prompting}
\bibfield{author}{\bibinfo{person}{Jonas Oppenlaender}, \bibinfo{person}{Rhema
  Linder}, {and} \bibinfo{person}{Johanna Silvennoinen}.}
  \bibinfo{year}{2024}\natexlab{}.
\newblock \showarticletitle{Prompting AI art: An investigation into the
  creative skill of prompt engineering}.
\newblock \bibinfo{journal}{\emph{International journal of human--computer
  interaction}} (\bibinfo{year}{2024}), \bibinfo{pages}{1--23}.
\newblock


\bibitem[Palmer et~al\mbox{.}(2008)]%
        {palmer2008aesthetic}
\bibfield{author}{\bibinfo{person}{Stephen~E Palmer},
  \bibinfo{person}{Jonathan~S Gardner}, {and} \bibinfo{person}{Thomas~D
  Wickens}.} \bibinfo{year}{2008}\natexlab{}.
\newblock \showarticletitle{Aesthetic issues in spatial composition: Effects of
  position and direction on framing single objects}.
\newblock \bibinfo{journal}{\emph{Spatial vision}} \bibinfo{volume}{21},
  \bibinfo{number}{3} (\bibinfo{year}{2008}), \bibinfo{pages}{421--450}.
\newblock


\bibitem[Palmer et~al\mbox{.}(2013)]%
        {palmer2013visual}
\bibfield{author}{\bibinfo{person}{Stephen~E Palmer}, \bibinfo{person}{Karen~B
  Schloss}, {and} \bibinfo{person}{Jonathan Sammartino}.}
  \bibinfo{year}{2013}\natexlab{}.
\newblock \showarticletitle{Visual aesthetics and human preference}.
\newblock \bibinfo{journal}{\emph{Annual review of psychology}}
  \bibinfo{volume}{64}, \bibinfo{number}{1} (\bibinfo{year}{2013}),
  \bibinfo{pages}{77--107}.
\newblock


\bibitem[Pan et~al\mbox{.}(2023)]%
        {pan2023drag}
\bibfield{author}{\bibinfo{person}{Xingang Pan}, \bibinfo{person}{Ayush
  Tewari}, \bibinfo{person}{Thomas Leimk{\"u}hler}, \bibinfo{person}{Lingjie
  Liu}, \bibinfo{person}{Abhimitra Meka}, {and} \bibinfo{person}{Christian
  Theobalt}.} \bibinfo{year}{2023}\natexlab{}.
\newblock \showarticletitle{Drag your gan: Interactive point-based manipulation
  on the generative image manifold}. In \bibinfo{booktitle}{\emph{ACM SIGGRAPH
  2023 conference proceedings}}. \bibinfo{pages}{1--11}.
\newblock


\bibitem[Poore(1976)]%
        {poore1976composition}
\bibfield{author}{\bibinfo{person}{Henry~Rankin Poore}.}
  \bibinfo{year}{1976}\natexlab{}.
\newblock \bibinfo{booktitle}{\emph{Composition in art}}.
\newblock \bibinfo{publisher}{Courier Corporation}.
\newblock


\bibitem[Qian et~al\mbox{.}(2022)]%
        {qian2022arnnotate}
\bibfield{author}{\bibinfo{person}{Xun Qian}, \bibinfo{person}{Fengming He},
  \bibinfo{person}{Xiyun Hu}, \bibinfo{person}{Tianyi Wang}, {and}
  \bibinfo{person}{Karthik Ramani}.} \bibinfo{year}{2022}\natexlab{}.
\newblock \showarticletitle{Arnnotate: An augmented reality interface for
  collecting custom dataset of 3d hand-object interaction pose estimation}. In
  \bibinfo{booktitle}{\emph{Proceedings of the 35th Annual ACM Symposium on
  User Interface Software and Technology}}. \bibinfo{pages}{1--14}.
\newblock


\bibitem[Qin et~al\mbox{.}(2023)]%
        {qin2023unicontrol}
\bibfield{author}{\bibinfo{person}{Can Qin}, \bibinfo{person}{Shu Zhang},
  \bibinfo{person}{Ning Yu}, \bibinfo{person}{Yihao Feng},
  \bibinfo{person}{Xinyi Yang}, \bibinfo{person}{Yingbo Zhou},
  \bibinfo{person}{Huan Wang}, \bibinfo{person}{Juan~Carlos Niebles},
  \bibinfo{person}{Caiming Xiong}, \bibinfo{person}{Silvio Savarese},
  {et~al\mbox{.}}} \bibinfo{year}{2023}\natexlab{}.
\newblock \showarticletitle{Unicontrol: A unified diffusion model for
  controllable visual generation in the wild}.
\newblock \bibinfo{journal}{\emph{arXiv preprint arXiv:2305.11147}}
  (\bibinfo{year}{2023}).
\newblock


\bibitem[Ramesh et~al\mbox{.}(2022)]%
        {ramesh2022hierarchical}
\bibfield{author}{\bibinfo{person}{Aditya Ramesh}, \bibinfo{person}{Prafulla
  Dhariwal}, \bibinfo{person}{Alex Nichol}, \bibinfo{person}{Casey Chu}, {and}
  \bibinfo{person}{Mark Chen}.} \bibinfo{year}{2022}\natexlab{}.
\newblock \showarticletitle{Hierarchical text-conditional image generation with
  clip latents}.
\newblock \bibinfo{journal}{\emph{arXiv preprint arXiv:2204.06125}}
  \bibinfo{volume}{1}, \bibinfo{number}{2} (\bibinfo{year}{2022}),
  \bibinfo{pages}{3}.
\newblock


\bibitem[Ramesh et~al\mbox{.}(2021)]%
        {ramesh2021zero}
\bibfield{author}{\bibinfo{person}{Aditya Ramesh}, \bibinfo{person}{Mikhail
  Pavlov}, \bibinfo{person}{Gabriel Goh}, \bibinfo{person}{Scott Gray},
  \bibinfo{person}{Chelsea Voss}, \bibinfo{person}{Alec Radford},
  \bibinfo{person}{Mark Chen}, {and} \bibinfo{person}{Ilya Sutskever}.}
  \bibinfo{year}{2021}\natexlab{}.
\newblock \showarticletitle{Zero-shot text-to-image generation}. In
  \bibinfo{booktitle}{\emph{International conference on machine learning}}.
  Pmlr, \bibinfo{pages}{8821--8831}.
\newblock


\bibitem[Reimers and Gurevych(2019)]%
        {reimers2019sentence}
\bibfield{author}{\bibinfo{person}{Nils Reimers} {and} \bibinfo{person}{Iryna
  Gurevych}.} \bibinfo{year}{2019}\natexlab{}.
\newblock \showarticletitle{Sentence-bert: Sentence embeddings using siamese
  bert-networks}.
\newblock \bibinfo{journal}{\emph{arXiv preprint arXiv:1908.10084}}
  (\bibinfo{year}{2019}).
\newblock


\bibitem[Reisman et~al\mbox{.}(2009)]%
        {reisman2009screen}
\bibfield{author}{\bibinfo{person}{Jason~L Reisman}, \bibinfo{person}{Philip~L
  Davidson}, {and} \bibinfo{person}{Jefferson~Y Han}.}
  \bibinfo{year}{2009}\natexlab{}.
\newblock \showarticletitle{A screen-space formulation for 2D and 3D direct
  manipulation}. In \bibinfo{booktitle}{\emph{Proceedings of the 22nd annual
  ACM symposium on User interface software and technology}}.
  \bibinfo{pages}{69--78}.
\newblock


\bibitem[Rombach et~al\mbox{.}(2022)]%
        {rombach2022high}
\bibfield{author}{\bibinfo{person}{Robin Rombach}, \bibinfo{person}{Andreas
  Blattmann}, \bibinfo{person}{Dominik Lorenz}, \bibinfo{person}{Patrick
  Esser}, {and} \bibinfo{person}{Bj{\"o}rn Ommer}.}
  \bibinfo{year}{2022}\natexlab{}.
\newblock \showarticletitle{High-resolution image synthesis with latent
  diffusion models}. In \bibinfo{booktitle}{\emph{Proceedings of the IEEE/CVF
  conference on computer vision and pattern recognition}}.
  \bibinfo{pages}{10684--10695}.
\newblock


\bibitem[Sammartino and Palmer(2012)]%
        {sammartino2012aesthetic}
\bibfield{author}{\bibinfo{person}{Jonathan Sammartino} {and}
  \bibinfo{person}{Stephen~E Palmer}.} \bibinfo{year}{2012}\natexlab{}.
\newblock \showarticletitle{Aesthetic issues in spatial composition: Effects of
  vertical position and perspective on framing single objects.}
\newblock \bibinfo{journal}{\emph{Journal of Experimental Psychology: Human
  Perception and Performance}} \bibinfo{volume}{38}, \bibinfo{number}{4}
  (\bibinfo{year}{2012}), \bibinfo{pages}{865}.
\newblock


\bibitem[Sample et~al\mbox{.}(2020)]%
        {sample2020components}
\bibfield{author}{\bibinfo{person}{Kevin~L Sample}, \bibinfo{person}{Henrik
  Hagtvedt}, {and} \bibinfo{person}{S~Adam Brasel}.}
  \bibinfo{year}{2020}\natexlab{}.
\newblock \showarticletitle{Components of visual perception in marketing
  contexts: A conceptual framework and review}.
\newblock \bibinfo{journal}{\emph{Journal of the Academy of Marketing Science}}
   \bibinfo{volume}{48} (\bibinfo{year}{2020}), \bibinfo{pages}{405--421}.
\newblock


\bibitem[Sarukkai et~al\mbox{.}(2024)]%
        {sarukkai2024block}
\bibfield{author}{\bibinfo{person}{Vishnu Sarukkai}, \bibinfo{person}{Lu Yuan},
  \bibinfo{person}{Mia Tang}, \bibinfo{person}{Maneesh Agrawala}, {and}
  \bibinfo{person}{Kayvon Fatahalian}.} \bibinfo{year}{2024}\natexlab{}.
\newblock \showarticletitle{Block and Detail: Scaffolding Sketch-to-Image
  Generation}. In \bibinfo{booktitle}{\emph{Proceedings of the 37th Annual ACM
  Symposium on User Interface Software and Technology}}.
  \bibinfo{pages}{1--13}.
\newblock


\bibitem[Schultheis et~al\mbox{.}(2012)]%
        {schultheis2012comparison}
\bibfield{author}{\bibinfo{person}{Udo Schultheis}, \bibinfo{person}{Jason
  Jerald}, \bibinfo{person}{Fernando Toledo}, \bibinfo{person}{Arun
  Yoganandan}, {and} \bibinfo{person}{Paul Mlyniec}.}
  \bibinfo{year}{2012}\natexlab{}.
\newblock \showarticletitle{Comparison of a two-handed interface to a wand
  interface and a mouse interface for fundamental 3D tasks}. In
  \bibinfo{booktitle}{\emph{2012 IEEE Symposium on 3D User Interfaces (3DUI)}}.
  IEEE, \bibinfo{pages}{117--124}.
\newblock


\bibitem[Shi et~al\mbox{.}(2025)]%
        {shi2025caring}
\bibfield{author}{\bibinfo{person}{Jingyu Shi}, \bibinfo{person}{Rahul Jain},
  \bibinfo{person}{Seungguen Chi}, \bibinfo{person}{Hyungjun Doh},
  \bibinfo{person}{Hyunggun Chi}, \bibinfo{person}{Alexander~J Quinn}, {and}
  \bibinfo{person}{Karthik Ramani}.} \bibinfo{year}{2025}\natexlab{}.
\newblock \showarticletitle{CARING-AI: Towards Authoring Context-aware
  Augmented Reality INstruction through Generative Artificial Intelligence}.
\newblock \bibinfo{journal}{\emph{arXiv preprint arXiv:2501.16557}}
  (\bibinfo{year}{2025}).
\newblock


\bibitem[Shi et~al\mbox{.}(2024a)]%
        {shi2024personalizing}
\bibfield{author}{\bibinfo{person}{Yang Shi}, \bibinfo{person}{Yechun Peng},
  \bibinfo{person}{Shengqi Dang}, \bibinfo{person}{Nanxuan Zhao}, {and}
  \bibinfo{person}{Nan Cao}.} \bibinfo{year}{2024}\natexlab{a}.
\newblock \showarticletitle{Personalizing Products with Stylized Head Portraits
  for Self-Expression}. In \bibinfo{booktitle}{\emph{Proceedings of the 2024
  CHI Conference on Human Factors in Computing Systems}}.
  \bibinfo{pages}{1--18}.
\newblock


\bibitem[Shi et~al\mbox{.}(2024b)]%
        {shi2024dragdiffusion}
\bibfield{author}{\bibinfo{person}{Yujun Shi}, \bibinfo{person}{Chuhui Xue},
  \bibinfo{person}{Jun~Hao Liew}, \bibinfo{person}{Jiachun Pan},
  \bibinfo{person}{Hanshu Yan}, \bibinfo{person}{Wenqing Zhang},
  \bibinfo{person}{Vincent~YF Tan}, {and} \bibinfo{person}{Song Bai}.}
  \bibinfo{year}{2024}\natexlab{b}.
\newblock \showarticletitle{Dragdiffusion: Harnessing diffusion models for
  interactive point-based image editing}. In
  \bibinfo{booktitle}{\emph{Proceedings of the IEEE/CVF Conference on Computer
  Vision and Pattern Recognition}}. \bibinfo{pages}{8839--8849}.
\newblock


\bibitem[Shin et~al\mbox{.}(2024)]%
        {shin2024instantdrag}
\bibfield{author}{\bibinfo{person}{Joonghyuk Shin}, \bibinfo{person}{Daehyeon
  Choi}, {and} \bibinfo{person}{Jaesik Park}.} \bibinfo{year}{2024}\natexlab{}.
\newblock \showarticletitle{InstantDrag: Improving Interactivity in Drag-based
  Image Editing}. In \bibinfo{booktitle}{\emph{SIGGRAPH Asia 2024 Conference
  Papers}}. \bibinfo{pages}{1--10}.
\newblock


\bibitem[Sikkel et~al\mbox{.}(2014)]%
        {sikkel2014clicking}
\bibfield{author}{\bibinfo{person}{Dirk Sikkel}, \bibinfo{person}{Reinder
  Steenbergen}, {and} \bibinfo{person}{Sjoerd Gras}.}
  \bibinfo{year}{2014}\natexlab{}.
\newblock \showarticletitle{Clicking vs. dragging: Different uses of the mouse
  and their implications for online surveys}.
\newblock \bibinfo{journal}{\emph{Public opinion quarterly}}
  \bibinfo{volume}{78}, \bibinfo{number}{1} (\bibinfo{year}{2014}),
  \bibinfo{pages}{177--190}.
\newblock


\bibitem[Sun et~al\mbox{.}(2024)]%
        {sun2024layoutvlm}
\bibfield{author}{\bibinfo{person}{Fan-Yun Sun}, \bibinfo{person}{Weiyu Liu},
  \bibinfo{person}{Siyi Gu}, \bibinfo{person}{Dylan Lim},
  \bibinfo{person}{Goutam Bhat}, \bibinfo{person}{Federico Tombari},
  \bibinfo{person}{Manling Li}, \bibinfo{person}{Nick Haber}, {and}
  \bibinfo{person}{Jiajun Wu}.} \bibinfo{year}{2024}\natexlab{}.
\newblock \showarticletitle{LayoutVLM: Differentiable Optimization of 3D Layout
  via Vision-Language Models}.
\newblock \bibinfo{journal}{\emph{arXiv preprint arXiv:2412.02193}}
  (\bibinfo{year}{2024}).
\newblock


\bibitem[Sun et~al\mbox{.}(2023)]%
        {sun2023dreamsync}
\bibfield{author}{\bibinfo{person}{Jiao Sun}, \bibinfo{person}{Deqing Fu},
  \bibinfo{person}{Yushi Hu}, \bibinfo{person}{Su Wang}, \bibinfo{person}{Royi
  Rassin}, \bibinfo{person}{Da-Cheng Juan}, \bibinfo{person}{Dana Alon},
  \bibinfo{person}{Charles Herrmann}, \bibinfo{person}{Sjoerd van Steenkiste},
  \bibinfo{person}{Ranjay Krishna}, {et~al\mbox{.}}}
  \bibinfo{year}{2023}\natexlab{}.
\newblock \showarticletitle{Dreamsync: Aligning text-to-image generation with
  image understanding feedback}.
\newblock \bibinfo{journal}{\emph{arXiv preprint arXiv:2311.17946}}
  (\bibinfo{year}{2023}).
\newblock


\bibitem[Svobodova et~al\mbox{.}(2014)]%
        {svobodova2014does}
\bibfield{author}{\bibinfo{person}{Kamila Svobodova}, \bibinfo{person}{Petr
  Sklenicka}, \bibinfo{person}{Kristina Molnarova}, {and} \bibinfo{person}{Jiri
  Vojar}.} \bibinfo{year}{2014}\natexlab{}.
\newblock \showarticletitle{Does the composition of landscape photographs
  affect visual preferences? The rule of the Golden Section and the position of
  the horizon}.
\newblock \bibinfo{journal}{\emph{Journal of Environmental Psychology}}
  \bibinfo{volume}{38} (\bibinfo{year}{2014}), \bibinfo{pages}{143--152}.
\newblock


\bibitem[Tevatia and Schaal(2000)]%
        {tevatia2000inverse}
\bibfield{author}{\bibinfo{person}{Gaurav Tevatia} {and}
  \bibinfo{person}{Stefan Schaal}.} \bibinfo{year}{2000}\natexlab{}.
\newblock \showarticletitle{Inverse kinematics for humanoid robots}. In
  \bibinfo{booktitle}{\emph{Proceedings 2000 ICRA. Millennium Conference. IEEE
  International Conference on Robotics and Automation. Symposia Proceedings
  (Cat. No. 00CH37065)}}, Vol.~\bibinfo{volume}{1}. IEEE,
  \bibinfo{pages}{294--299}.
\newblock


\bibitem[{Unity Technologies}({[n.\,d.]})]%
        {UnityWebsite}
\bibfield{author}{\bibinfo{person}{{Unity Technologies}}.}
  \bibinfo{year}{[n.\,d.]}\natexlab{}.
\newblock \bibinfo{title}{Unity Real-Time Development Platform | 3D, 2D, VR \&
  AR Engine}.
\newblock
\urldef\tempurl%
\url{https://unity.com/}
\showURL{%
\tempurl}
\newblock
\shownote{Accessed: 2025-04-08}.


\bibitem[Visser et~al\mbox{.}(1999)]%
        {visser1999attentional}
\bibfield{author}{\bibinfo{person}{Troy~AW Visser}, \bibinfo{person}{Walter~F
  Bischof}, {and} \bibinfo{person}{Vincent Di~Lollo}.}
  \bibinfo{year}{1999}\natexlab{}.
\newblock \showarticletitle{Attentional switching in spatial and nonspatial
  domains: Evidence from the attentional blink.}
\newblock \bibinfo{journal}{\emph{Psychological Bulletin}}
  \bibinfo{volume}{125}, \bibinfo{number}{4} (\bibinfo{year}{1999}),
  \bibinfo{pages}{458}.
\newblock


\bibitem[Wadinambiarachchi et~al\mbox{.}(2024)]%
        {wadinambiarachchi2024effects}
\bibfield{author}{\bibinfo{person}{Samangi Wadinambiarachchi},
  \bibinfo{person}{Ryan~M Kelly}, \bibinfo{person}{Saumya Pareek},
  \bibinfo{person}{Qiushi Zhou}, {and} \bibinfo{person}{Eduardo Velloso}.}
  \bibinfo{year}{2024}\natexlab{}.
\newblock \showarticletitle{The effects of generative ai on design fixation and
  divergent thinking}. In \bibinfo{booktitle}{\emph{Proceedings of the 2024 CHI
  Conference on Human Factors in Computing Systems}}. \bibinfo{pages}{1--18}.
\newblock


\bibitem[Wang et~al\mbox{.}(2024b)]%
        {wang2024roomdreaming}
\bibfield{author}{\bibinfo{person}{Shun-Yu Wang}, \bibinfo{person}{Wei-Chung
  Su}, \bibinfo{person}{Serena Chen}, \bibinfo{person}{Ching-Yi Tsai},
  \bibinfo{person}{Marta Misztal}, \bibinfo{person}{Katherine~M Cheng},
  \bibinfo{person}{Alwena Lin}, \bibinfo{person}{Yu Chen}, {and}
  \bibinfo{person}{Mike~Y Chen}.} \bibinfo{year}{2024}\natexlab{b}.
\newblock \showarticletitle{Roomdreaming: Generative-AI approach to
  facilitating iterative, preliminary interior design exploration}. In
  \bibinfo{booktitle}{\emph{Proceedings of the 2024 CHI Conference on Human
  Factors in Computing Systems}}. \bibinfo{pages}{1--20}.
\newblock


\bibitem[Wang et~al\mbox{.}(2021)]%
        {wang2021gesturar}
\bibfield{author}{\bibinfo{person}{Tianyi Wang}, \bibinfo{person}{Xun Qian},
  \bibinfo{person}{Fengming He}, \bibinfo{person}{Xiyun Hu},
  \bibinfo{person}{Yuanzhi Cao}, {and} \bibinfo{person}{Karthik Ramani}.}
  \bibinfo{year}{2021}\natexlab{}.
\newblock \showarticletitle{Gesturar: An authoring system for creating freehand
  interactive augmented reality applications}. In \bibinfo{booktitle}{\emph{The
  34th Annual ACM Symposium on User Interface Software and Technology}}.
  \bibinfo{pages}{552--567}.
\newblock


\bibitem[Wang et~al\mbox{.}(2024a)]%
        {wang2024instancediffusion}
\bibfield{author}{\bibinfo{person}{Xudong Wang}, \bibinfo{person}{Trevor
  Darrell}, \bibinfo{person}{Sai~Saketh Rambhatla}, \bibinfo{person}{Rohit
  Girdhar}, {and} \bibinfo{person}{Ishan Misra}.}
  \bibinfo{year}{2024}\natexlab{a}.
\newblock \showarticletitle{Instancediffusion: Instance-level control for image
  generation}. In \bibinfo{booktitle}{\emph{Proceedings of the IEEE/CVF
  Conference on Computer Vision and Pattern Recognition}}.
  \bibinfo{pages}{6232--6242}.
\newblock


\bibitem[Wang et~al\mbox{.}(2023)]%
        {wang2023language}
\bibfield{author}{\bibinfo{person}{Zhenwei Wang}, \bibinfo{person}{Nanxuan
  Zhao}, \bibinfo{person}{Gerhard~P Hancke}, {and} \bibinfo{person}{Rynson~WH
  Lau}.} \bibinfo{year}{2023}\natexlab{}.
\newblock \showarticletitle{Language-based Photo Color Adjustment for Graphic
  Designs.}
\newblock \bibinfo{journal}{\emph{ACM Trans. Graph.}} \bibinfo{volume}{42},
  \bibinfo{number}{4} (\bibinfo{year}{2023}), \bibinfo{pages}{101--1}.
\newblock


\bibitem[Wei et~al\mbox{.}(2023)]%
        {wei2023lego}
\bibfield{author}{\bibinfo{person}{Qiuhong~Anna Wei}, \bibinfo{person}{Sijie
  Ding}, \bibinfo{person}{Jeong~Joon Park}, \bibinfo{person}{Rahul Sajnani},
  \bibinfo{person}{Adrien Poulenard}, \bibinfo{person}{Srinath Sridhar}, {and}
  \bibinfo{person}{Leonidas Guibas}.} \bibinfo{year}{2023}\natexlab{}.
\newblock \showarticletitle{Lego-net: Learning regular rearrangements of
  objects in rooms}. In \bibinfo{booktitle}{\emph{Proceedings of the IEEE/CVF
  Conference on Computer Vision and Pattern Recognition}}.
  \bibinfo{pages}{19037--19047}.
\newblock


\bibitem[Williford et~al\mbox{.}(2019)]%
        {williford2019framework}
\bibfield{author}{\bibinfo{person}{Blake Williford}, \bibinfo{person}{Matthew
  Runyon}, \bibinfo{person}{Josh Cherian}, \bibinfo{person}{Wayne Li},
  \bibinfo{person}{Julie Linsey}, {and} \bibinfo{person}{Tracy Hammond}.}
  \bibinfo{year}{2019}\natexlab{}.
\newblock \showarticletitle{A framework for motivating sketching practice with
  sketch-based gameplay}. In \bibinfo{booktitle}{\emph{Proceedings of the
  Annual Symposium on Computer-Human Interaction in Play}}.
  \bibinfo{pages}{533--544}.
\newblock


\bibitem[Wu et~al\mbox{.}(2015)]%
        {wu2015touchsketch}
\bibfield{author}{\bibinfo{person}{Siju Wu}, \bibinfo{person}{Amine Chellali},
  \bibinfo{person}{Samir Otmane}, {and} \bibinfo{person}{Guillaume Moreau}.}
  \bibinfo{year}{2015}\natexlab{}.
\newblock \showarticletitle{TouchSketch: a touch-based interface for 3D object
  manipulation and editing}. In \bibinfo{booktitle}{\emph{Proceedings of the
  21st ACM Symposium on Virtual Reality Software and Technology}}.
  \bibinfo{pages}{59--68}.
\newblock


\bibitem[Yang et~al\mbox{.}(2024)]%
        {yang2024emogen}
\bibfield{author}{\bibinfo{person}{Jingyuan Yang}, \bibinfo{person}{Jiawei
  Feng}, {and} \bibinfo{person}{Hui Huang}.} \bibinfo{year}{2024}\natexlab{}.
\newblock \showarticletitle{EmoGen: Emotional image content generation with
  text-to-image diffusion models}. In \bibinfo{booktitle}{\emph{Proceedings of
  the IEEE/CVF Conference on Computer Vision and Pattern Recognition}}.
  \bibinfo{pages}{6358--6368}.
\newblock


\bibitem[Yang et~al\mbox{.}(2019)]%
        {yang2019aesthetic}
\bibfield{author}{\bibinfo{person}{Taoxi Yang}, \bibinfo{person}{Sarita
  Silveira}, \bibinfo{person}{Arusu Formuli}, \bibinfo{person}{Marco Paolini},
  \bibinfo{person}{Ernst P{\"o}ppel}, \bibinfo{person}{Tilmann Sander}, {and}
  \bibinfo{person}{Yan Bao}.} \bibinfo{year}{2019}\natexlab{}.
\newblock \showarticletitle{Aesthetic experiences across cultures: Neural
  correlates when viewing traditional Eastern or Western landscape paintings}.
\newblock \bibinfo{journal}{\emph{Frontiers in psychology}}
  \bibinfo{volume}{10} (\bibinfo{year}{2019}), \bibinfo{pages}{798}.
\newblock


\bibitem[Zeng et~al\mbox{.}(2024)]%
        {zeng2024paint3d}
\bibfield{author}{\bibinfo{person}{Xianfang Zeng}, \bibinfo{person}{Xin Chen},
  \bibinfo{person}{Zhongqi Qi}, \bibinfo{person}{Wen Liu},
  \bibinfo{person}{Zibo Zhao}, \bibinfo{person}{Zhibin Wang},
  \bibinfo{person}{Bin Fu}, \bibinfo{person}{Yong Liu}, {and}
  \bibinfo{person}{Gang Yu}.} \bibinfo{year}{2024}\natexlab{}.
\newblock \showarticletitle{Paint3d: Paint anything 3d with lighting-less
  texture diffusion models}. In \bibinfo{booktitle}{\emph{Proceedings of the
  IEEE/CVF conference on computer vision and pattern recognition}}.
  \bibinfo{pages}{4252--4262}.
\newblock


\bibitem[Zhang et~al\mbox{.}(2024)]%
        {zhang2024protodreamer}
\bibfield{author}{\bibinfo{person}{Hongbo Zhang}, \bibinfo{person}{Pei Chen},
  \bibinfo{person}{Xuelong Xie}, \bibinfo{person}{Chaoyi Lin},
  \bibinfo{person}{Lianyan Liu}, \bibinfo{person}{Zhuoshu Li},
  \bibinfo{person}{Weitao You}, {and} \bibinfo{person}{Lingyun Sun}.}
  \bibinfo{year}{2024}\natexlab{}.
\newblock \showarticletitle{Protodreamer: A mixed-prototype tool combining
  physical model and generative AI to support conceptual design}. In
  \bibinfo{booktitle}{\emph{Proceedings of the 37th Annual ACM Symposium on
  User Interface Software and Technology}}. \bibinfo{pages}{1--18}.
\newblock


\bibitem[Zhang et~al\mbox{.}(2023a)]%
        {zhang2023adding}
\bibfield{author}{\bibinfo{person}{Lvmin Zhang}, \bibinfo{person}{Anyi Rao},
  {and} \bibinfo{person}{Maneesh Agrawala}.} \bibinfo{year}{2023}\natexlab{a}.
\newblock \showarticletitle{Adding conditional control to text-to-image
  diffusion models}. In \bibinfo{booktitle}{\emph{Proceedings of the IEEE/CVF
  international conference on computer vision}}. \bibinfo{pages}{3836--3847}.
\newblock


\bibitem[Zhang et~al\mbox{.}(2025)]%
        {zhang2025scaling}
\bibfield{author}{\bibinfo{person}{Lvmin Zhang}, \bibinfo{person}{Anyi Rao},
  {and} \bibinfo{person}{Maneesh Agrawala}.} \bibinfo{year}{2025}\natexlab{}.
\newblock \showarticletitle{Scaling in-the-wild training for diffusion-based
  illumination harmonization and editing by imposing consistent light
  transport}. In \bibinfo{booktitle}{\emph{The Thirteenth International
  Conference on Learning Representations}}.
\newblock


\bibitem[Zhang et~al\mbox{.}(2023b)]%
        {zhang2023controllable}
\bibfield{author}{\bibinfo{person}{Tianjun Zhang}, \bibinfo{person}{Yi Zhang},
  \bibinfo{person}{Vibhav Vineet}, \bibinfo{person}{Neel Joshi}, {and}
  \bibinfo{person}{Xin Wang}.} \bibinfo{year}{2023}\natexlab{b}.
\newblock \showarticletitle{Controllable text-to-image generation with gpt-4}.
\newblock \bibinfo{journal}{\emph{arXiv preprint arXiv:2305.18583}}
  (\bibinfo{year}{2023}).
\newblock


\bibitem[Zhang et~al\mbox{.}(2019)]%
        {zhang2019facing}
\bibfield{author}{\bibinfo{person}{Yuli Zhang}, \bibinfo{person}{Hyokjin Kwak},
  \bibinfo{person}{Haeyoung Jeong}, {and} \bibinfo{person}{Marina Puzakova}.}
  \bibinfo{year}{2019}\natexlab{}.
\newblock \showarticletitle{Facing the “right” side? The effect of product
  facing direction}.
\newblock \bibinfo{journal}{\emph{Journal of Advertising}}
  \bibinfo{volume}{48}, \bibinfo{number}{2} (\bibinfo{year}{2019}),
  \bibinfo{pages}{153--166}.
\newblock


\bibitem[Zhang et~al\mbox{.}(2012)]%
        {zhang2012aesthetic}
\bibfield{author}{\bibinfo{person}{Yanhao Zhang}, \bibinfo{person}{Xiaoshuai
  Sun}, \bibinfo{person}{Hongxun Yao}, \bibinfo{person}{Lei Qin}, {and}
  \bibinfo{person}{Qingming Huang}.} \bibinfo{year}{2012}\natexlab{}.
\newblock \showarticletitle{Aesthetic composition represetation for portrait
  photographing recommendation}. In \bibinfo{booktitle}{\emph{2012 19th IEEE
  international conference on image processing}}. IEEE,
  \bibinfo{pages}{2753--2756}.
\newblock


\bibitem[Zhao et~al\mbox{.}(2025)]%
        {zhao2025hunyuan3d}
\bibfield{author}{\bibinfo{person}{Zibo Zhao}, \bibinfo{person}{Zeqiang Lai},
  \bibinfo{person}{Qingxiang Lin}, \bibinfo{person}{Yunfei Zhao},
  \bibinfo{person}{Haolin Liu}, \bibinfo{person}{Shuhui Yang},
  \bibinfo{person}{Yifei Feng}, \bibinfo{person}{Mingxin Yang},
  \bibinfo{person}{Sheng Zhang}, \bibinfo{person}{Xianghui Yang},
  {et~al\mbox{.}}} \bibinfo{year}{2025}\natexlab{}.
\newblock \showarticletitle{Hunyuan3d 2.0: Scaling diffusion models for high
  resolution textured 3d assets generation}.
\newblock \bibinfo{journal}{\emph{arXiv preprint arXiv:2501.12202}}
  (\bibinfo{year}{2025}).
\newblock


\bibitem[Zheng et~al\mbox{.}(2023)]%
        {zheng2023layoutdiffusion}
\bibfield{author}{\bibinfo{person}{Guangcong Zheng}, \bibinfo{person}{Xianpan
  Zhou}, \bibinfo{person}{Xuewei Li}, \bibinfo{person}{Zhongang Qi},
  \bibinfo{person}{Ying Shan}, {and} \bibinfo{person}{Xi Li}.}
  \bibinfo{year}{2023}\natexlab{}.
\newblock \showarticletitle{Layoutdiffusion: Controllable diffusion model for
  layout-to-image generation}. In \bibinfo{booktitle}{\emph{Proceedings of the
  IEEE/CVF Conference on Computer Vision and Pattern Recognition}}.
  \bibinfo{pages}{22490--22499}.
\newblock


\bibitem[Zhou et~al\mbox{.}(2022)]%
        {zhou2022simple}
\bibfield{author}{\bibinfo{person}{Xingyi Zhou}, \bibinfo{person}{Vladlen
  Koltun}, {and} \bibinfo{person}{Philipp Kr{\"a}henb{\"u}hl}.}
  \bibinfo{year}{2022}\natexlab{}.
\newblock \showarticletitle{Simple multi-dataset detection}. In
  \bibinfo{booktitle}{\emph{Proceedings of the IEEE/CVF conference on computer
  vision and pattern recognition}}. \bibinfo{pages}{7571--7580}.
\newblock


\bibitem[Zhu et~al\mbox{.}(2022)]%
        {zhu2022mecharspace}
\bibfield{author}{\bibinfo{person}{Zhengzhe Zhu}, \bibinfo{person}{Ziyi Liu},
  \bibinfo{person}{Tianyi Wang}, \bibinfo{person}{Youyou Zhang},
  \bibinfo{person}{Xun Qian}, \bibinfo{person}{Pashin~Farsak Raja},
  \bibinfo{person}{Ana Villanueva}, {and} \bibinfo{person}{Karthik Ramani}.}
  \bibinfo{year}{2022}\natexlab{}.
\newblock \showarticletitle{MechARspace: An authoring system enabling
  bidirectional binding of augmented reality with toys in real-time}. In
  \bibinfo{booktitle}{\emph{Proceedings of the 35th Annual ACM Symposium on
  User Interface Software and Technology}}. \bibinfo{pages}{1--16}.
\newblock


\end{thebibliography}

\appendix

\section{Appendix}

\subsection{Baseline System Interface}

\begin{figure}[h]
  \centering
  \includegraphics[width=\linewidth]{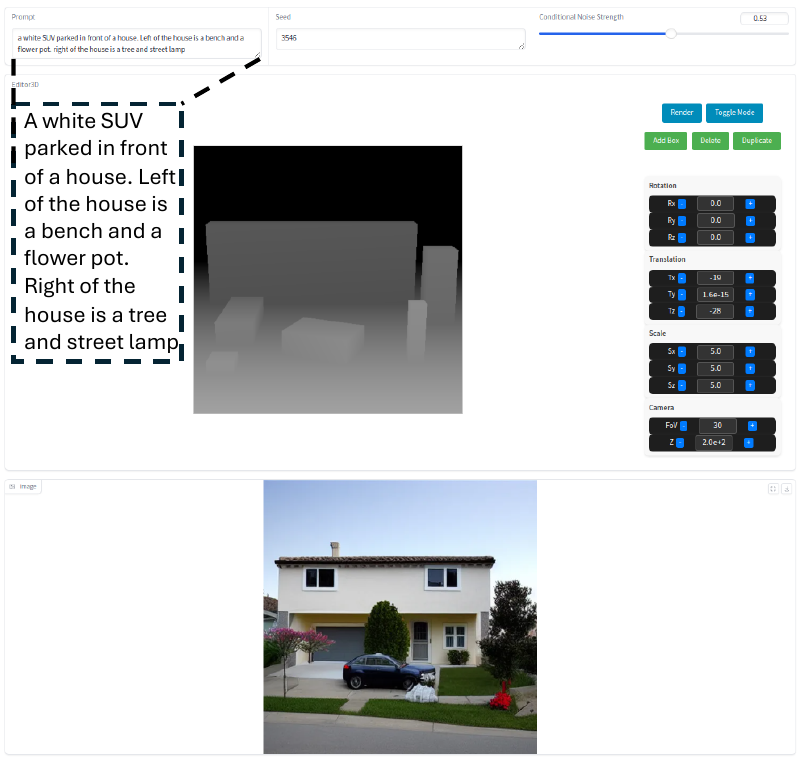}
  \caption{Baseline Interface}
  \Description{None}
  \label{fig:baseline interface}
\end{figure}

\subsection{System Usability Scale}
\begin{figure}[htp]
    \centering
    \includegraphics[width=0.5\textwidth]{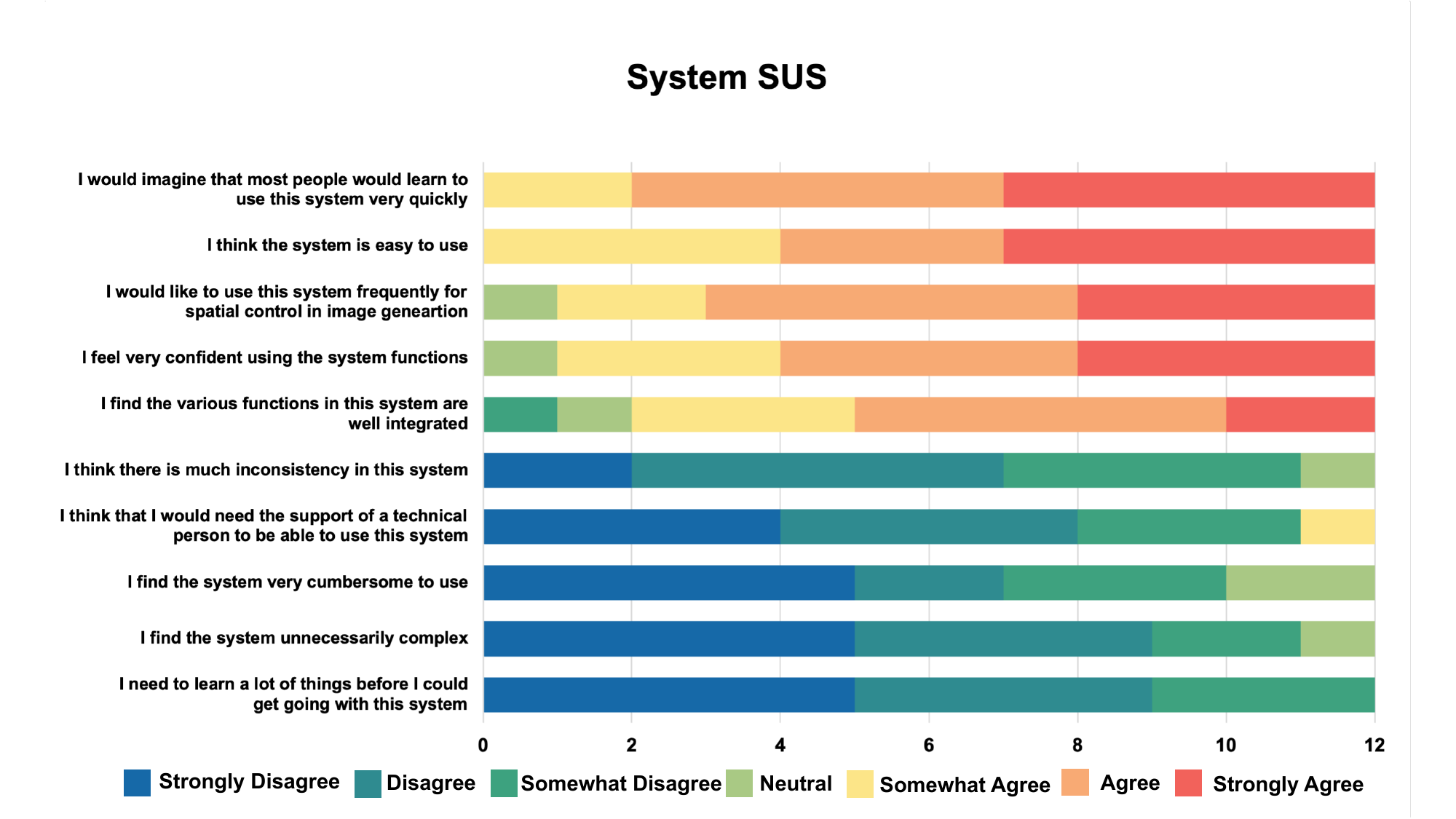}
    \caption{System Usability Scale}
    \label{fig:my_label}
\end{figure}

\subsubsection{Datasets}

We manually selected 20 common indoor and 20 common outdoor object categories from the ShapeNet~\cite{chang2015shapenet} and Objaverse~\cite{deitke2023objaverse} datasets. Most of these 3D models are collected from ShapeNet; however, for categories containing fewer than 300 models, we supplemented them with additional models from Objaverse. The following table lists the selected categories and the corresponding number of models in each category:

\begin{table}[htp]
\centering
\begin{tabular}{l r | l r}
\hline
\textbf{Indoor Categories} & \textbf{Count} & \textbf{Outdoor Categories} & \textbf{Count} \\
\hline
table      & 8443  & table             & 8443   \\
chair      & 6778  & chair             & 6778   \\
sofa       & 3173  & lamp              & 2318   \\
lamp       & 2318   & watercraft  & 1939    \\
cabinet    & 1572  & bench             & 1816   \\
display    & 1193  & motorbike         & 337   \\
telephone  & 1052  & bus               & 939    \\
guitar     & 787  & flowerpot         & 602    \\
clock      & 655   & trash bin         & 374    \\
flowerpot  & 602   & building            & 300    \\
jar        & 597   & car               & 300    \\
bottle     & 590   & bicycle           & 300    \\
bookshelf  & 466   & skateboard        & 300    \\
laptop     & 450   & mailbox           & 300    \\
trash bin  & 374   & bushes            & 300    \\
pillow     & 300   & birdhouse         & 300    \\
bag        & 300   & pier             & 300    \\
basket     & 300   & fountain          & 300    \\
bed        & 300   & flower             & 300    \\
mug        & 300   & sculpture         & 300    \\
\hline
\end{tabular}
\caption{Indoor and Outdoor Object Cateogries and Counts}
\end{table}

\subsubsection{Meta Prompt for Object Registration}
\textbf{System Content:}\\
You are a professional scene designer. You have 20 [indoor/outdoor] furniture categories (listed as [categories list]). Based on the user's requirements, select the categories needed to create the requested scene. For each chosen category, specify how many items from that category should be included. You should also consider the functional interdependencies among items. For instance, a laptop necessitates a table to place; a pillow implies a bed in the scene.

Provide your answer in the following format:
\begin{center}
CATEGORY\_NAME\_1: 2, \\
CATEGORY\_NAME\_2: 1
\end{center}

\noindent
\textbf{User Content:}
\begin{align*}
\text{User Prompt:} &\quad \text{[user\_text\_prompt]} \\
\end{align*}

\subsubsection{Meta Prompt for Scene Synthesis}
\textbf{System Content:}\\
Scene design in 300×300 coordinate system

Scene specifications:

1. Size: 300×300 units

2. Coordinate system: Origin (0, 0) is located in the upper left corner; (300, 300) is located in the lower right corner

3. The x-axis spans the scene's width (0 to 300), The y-axis spans the scene's height (0 to 300)

Placement Goals:

1. Each object can be simplified to a rectangle in a 2D top view, with the dimensions specified on the "Item size dictionary" part, specifically in (length of the front, length of the side) order. Its "front direction" is the primary orientation of the object during normal use. Its "front face" is the edge that aligns with the "front direction". Its "side face" is the edge perprndicular to the "front face".

2. Grounded objects are defined as items stand directly on the floor (e.g., tables); accessory objects are objects placed on top of these grounded objects (e.g., mugs). Place the grounded objects first, then place the accessory objects based on their results. 

3. Place items in a practical and aesthetically pleasing way (e.g., chairs near a table, plants near a corner, etc.).

4. Make sure all coordinates stay within the 300×300 boundary.

5. For items with multiple instances (e.g., 2 chairs), clearly label them (e.g., Chair 1, Chair 2).

Task:

1. Based on the above requirements, come up with the center coordinates and rotation angle (i.e., the counter-clockwise angle between their front direction and the standard negative y-axis, typically from [0, 90, 180, 270]) for each item listed. 

 
2. Based on the location result, list all the spatial affordance up and down relationships in the result, and save the result in the form of (top, down). (e.g., (bottle 1, desk) means that the first bottle is placed on the only desk).

Output format:

The center coordinates, orientation of each item and their relationship are returned as:

\noindent
\text{Location:}
\begin{align*}
\text{CATEGORY\_NAME\_1:} &\quad (x_1, y_1, \theta_1) \\
\text{CATEGORY\_NAME\_2:} &\quad (x_2, y_2, \theta_2)
\end{align*}
\noindent
\text{Relation:}
\begin{align*}
&[(\text{CATEGORY\_NAME\_1}, \text{CATEGORY\_NAME\_2})]
\end{align*}

\noindent
\textbf{User Content:}
\begin{align*}
\text{User Prompt:} &\quad \text{[user\_text\_prompt]} \\
\text{Items to Place:} &\quad \text{[item\_dict]} \\
\text{Item Size (front face, side face):} &\quad \text{[item\_size\_dict]} \\
\end{align*}
\subsubsection{Caption generation in GPT-CLIP Score}
\textbf{Input Text:}\\
Compose a concise (around 40 words) caption that accurately lists every visible object in the image and briefly describes each object’s appearance, prioritizing clarity, completeness, and correctness.

\noindent
\textbf{Input Image:}
\begin{align*}
\text{Input Image:} &\quad \text{[b64encode\_image]}
\end{align*}
\subsubsection{GPT-4V Spatial Score in Metrics}
\textbf{Input Text:}\\

You are my assistant to identify objects and their spatial layout in the image. \
                According to the image, evaluate if the text [prompt] is correctly portrayed in the image. \
                Give a score from 0 to 100, according the criteria: \\\
                100: correct spatial layout in the image for all objects mentioned in the text. \\
                80: basically, spatial layout of objects matches the text. \\
                60: spatial layout not aligned properly with the text. \\
                40: image not aligned properly with the text. \\
                20: image almost irrelevant to the text. \\
                Provide your analysis and explanation in JSON format with the following keys: score (e.g., 85), \
                explanation (within 20 words).

\noindent
\textbf{Input Image:}
\begin{align*}
\text{Input Image:} &\quad \text{[b64encode\_image]}
\end{align*}

\subsection{Target Image for Controllabily Evaluation}
\begin{figure}[htp]
    \centering
    \includegraphics[width=0.4\textwidth,height=5.5cm, keepaspectratio=false]{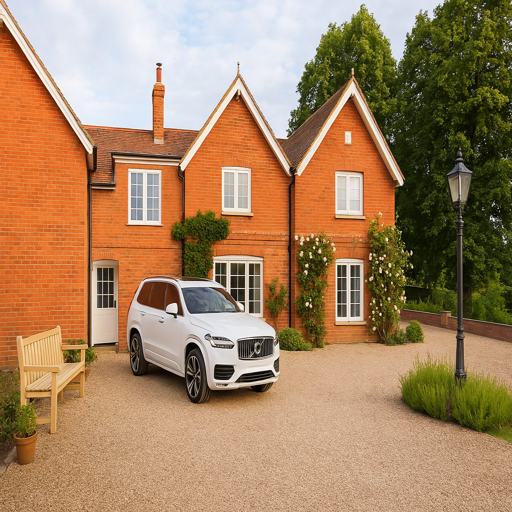}
    \caption{Target Image for Controllabily Evaluation}
    \label{fig:Target Imagel}
\end{figure}

\begin{figure}[htp]
    \centering
    \includegraphics[width=0.4\textwidth,height=5.5cm, keepaspectratio=false]{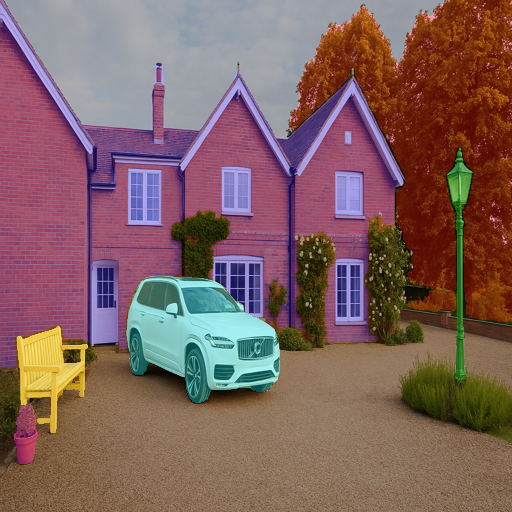}
    \caption{Masked Target Image for Controllabily Evaluation}
    \label{fig:Masked Target Image}
\end{figure}

\subsection{Count and Time in Close-Ended Experiment}

\begin{table}[htbp]
\centering
\caption{Count and Time in Controllabily Evaluation}
\label{tab:performance_metrics}
\begin{tabular}{l | cc | cc | cc} 
  \toprule
  & \multicolumn{2}{c|}{\textbf{Canvas3D}} 
  & \multicolumn{2}{c|}{\textbf{Baseline}} 
  & \multicolumn{2}{c}{\textbf{Statistics}} \\
& \textbf{Mean} & \textbf{Std} 
                      & \textbf{Mean} & \textbf{Std} 
                      & \textbf{p}    & \textbf{Sig} \\
  \midrule
  \# of generated img & 6.08 & 3.75  & 7.17 & 3.88   & 0.25 &  \\
  \# of liked img     & 1.75   & 0.87   & 1.0    & 1.81   & 0.5 & \\
  liked ratio          & 0.34 & 0.16  & 0.11 & 0.15   & 0.25 & \\
  time(s) to first liked      & 448.0  & 183.52& 786.8  & 374.01 &  & \\
  \bottomrule
\end{tabular}
\end{table}

\subsection{Count and Time in Open-Ended Experiment}

\begin{table}[htp]
\centering
\caption{Count and Time in Usability Evalution}
\label{tab:example}
\begin{tabular}{l | cc | cc} 
  \toprule
  & \multicolumn{2}{c|}{\textbf{Indoor Scene}} 
  & \multicolumn{2}{c}{\textbf{Human Pose}} \\
  & \textbf{Mean} & \textbf{Std} 
  & \textbf{Mean} & \textbf{Std} \\
  \midrule
  \# of generated img & 4.0    & 1.33 
                        & 9.33 & 3.94 \\
  \# of liked img     & 1.40    & 0.70 
                        & 1.58 & 1.16 \\
                        
  liked ratio           & 0.39  & 0.22 
                        & 0.20 & 0.15 \\
  time(s) to first liked     & 239.56 & 132.84 
                        & 388.20  & 246.84 \\
  \bottomrule
\end{tabular}
\end{table}


\subsection{Brief Workflow of Quality Metrics}

\begin{itemize}
    \item \textbf{CLIP Score}: Calculate the CLIP score between the user's input prompt and the generated image.
    \item {\textbf{GPT Spatial Score}}: Use a meta-prompt to instruct GPT to act as a judge that evaluates the spatial similarity between the input text and the generated image.
    \item {\textbf{GPT-CLIP Score}}: Let ChatGPT generate a caption for the target image, denoted as target\_caption. Then, using the same prompt and ChatGPT model, generate captions for the user-generated images, and calculate the CLIP score between this caption with target\_caption. This algorithm is similar to B-CLIP, but substitutes the BLIP model with ChatGPT.
    \item {\textbf{Uni-Det Score}}: Given 5 pre-defined 3D spatial relationships among objects within the Intended Object Set. 
    For each specified relationship, the system uses UniDet \cite{zhou2022simple} to detect objects in both the target image and the generated image. Then, it calculates a score comparing the positions and depths of corresponding object bounding boxes between the two images. The average of those five scores is the overall score.
    \item {\textbf{Recall Score}}: Specify a intended object set (6 objects in total, refer masked target image) and calculate the recall rate of this object within the generated image.
    
\end{itemize}

The 3D spatial relationship defined for the calculation of the Uni-Det score.

\begin{itemize}
    \item 'house', 'at the front left of', 'trees'
    \item 'house', 'at the back left of', 'lamp'
    \item 'house', 'at the back of', 'car'
    \item 'car', 'at the back right of', 'bench'
    \item 'bench', 'at the back of', 'flowerpot'
\end{itemize}

Intended Object Set for Recall Score:

\begin{itemize}
    \item House
    \item Trees
    \item Lamp
    \item Car
    \item Bench
    \item Flowerpot
\end{itemize}

\subsection{User Study Demographic}

\begin{table*}[!htp]
\centering
\footnotesize
\begin{tabular}{|c|c|c|c|c|c|}
\hline
User ID & Gender & Educational Level & Academic Discipline or Major & Image Generation Software Used & Image Generation Frequency (Monthly) \\
\hline
User 1 & Male   & Graduate   & Statistics  &  & Rarely (1-5 times) \\
\hline
User 2 & Male   & Graduate   & Public Health  &  & Never \\
\hline
User 3 & Male   & Graduate   & Robotics  &  & Never \\
\hline
User 4 & Female & Graduate   & Human Computer Interaction  & DallE, ChatGPT  & Occasionally (5-10 times) \\
\hline
User 5 & Male   & Graduate   & Mechanical Engineering  & Midjourney  & Rarely (1-5 times) \\
\hline
User 6 & Male   & Undergraduate       & Computer Information Systems  &  & Rarely (1-5 times) \\
\hline
User 7$^\ast$ & Male   & Undergraduate       & Game Design  & Photoshop AI, Microsoft AI  & Occasionally (5-10 times) \\
\hline
User 8$^\ast$ & Female & Undergraduate       & Visual Effects and Animation  & DallE  & Rarely (1-5 times) \\
\hline
User 9$^\ast$ & Male   & Undergraduate       & Computer Graphics, Game Design   & DallE, ChatGPT  & Frequently (10-20 times) \\
\hline
User 10$^\ast$ & Male  & Undergraduate       & User Experience Design, Game Design  & ChatGPT, Midjourney  & Frequently (10-20 times) \\
\hline
User 11 & Female & Graduate   & Mechanical Engineering  & ChatGPT  & Rarely (1-5 times) \\
\hline
User 12$^\ast$ & Male  & Graduate   & Commercial Art  & Stable Diffusion, Novel AI  & Very Frequently (more than 20 times) \\
\hline
\end{tabular}
\caption{User Study Demographic ($\ast$ denotes expert)}
\label{tab:user_responses}
\end{table*}

\end{document}